\documentclass[sn-mathphys,Numbered,iicol]{sn-jnl}
\usepackage{graphicx}%
\usepackage{multirow}%
\usepackage{amsmath,amssymb,amsfonts}%
\usepackage{amsthm}%
\usepackage{mathrsfs}%
\usepackage[title]{appendix}%
\usepackage{xcolor}%
\usepackage{textcomp}%
\usepackage{manyfoot}%
\usepackage{booktabs}%
\usepackage{algorithm}%
\usepackage{algorithmicx}%
\usepackage{algpseudocode}%
\usepackage{listings}%

\usepackage[numbers]{natbib}
\usepackage{graphicx} 
\usepackage{float} 
\usepackage{subfigure} 
\usepackage{amsmath}
\usepackage{bm}
\usepackage[justification=centering]{caption}
\usepackage{multirow}
\usepackage{float}

\theoremstyle{thmstyleone}%
\theoremstyle{thmstyletwo}%

\theoremstyle{thmstylethree}%

\begin{document}

\title{Multi-Omics Fusion with Soft Labeling for Enhanced Prediction of Distant Metastasis in Nasopharyngeal Carcinoma Patients after Radiotherapy}

\author[1,2]{\fnm{Jiabao} \sur{SHENG}}\email{jia-bao.sheng@connect.polyu.hk}

\author[2,3]{\fnm{SaiKit} \sur{LAM}}\email{saikit.lam@polyu.edu.hk}

\author[1]{\fnm{Jiang} \sur{ZHANG}}\email{jiang.zhang@connect.polyu.hk}

\author[1]{\fnm{Yuanpeng} \sur{ZHANG}}\email{y.p.zhang@ieee.org}

\author*[1,2]{\fnm{Jing} \sur{CAI}}\email{jing.cai@polyu.edu.hk}

\affil[1]{\orgdiv{Department of Health Technology and Informatics}, \orgname{The Hong Kong Polytechnic University}, \city{Hong Kong SAR}}

\affil[2]{\orgdiv{Research Institute for Smart Ageing}, \orgname{The Hong Kong Polytechnic University}, \city{Hong Kong SAR}}

\affil[3]{\orgdiv{Department of Biomedical Engineering}, \orgname{The Hong Kong Polytechnic University}, \city{Hong Kong SAR}}


\abstract{
Omics fusion has emerged as a crucial preprocessing approach in the field of medical image processing, providing significant assistance to several studies. One of the challenges encountered in the integration of omics data is the presence of unpredictability arising from disparities in data sources and medical imaging equipment. Due to the presence of these differences, the distribution of omics futures exhibits spatial heterogeneity, hence diminishing their capacity to enhance subsequent tasks. In order to overcome this challenge and facilitate the integration of their joint application to specific medical objectives, this study aims to develop a fusion methodology that mitigates the disparities inherent in omics data. The utilization of the multi-kernel late-fusion method has gained significant popularity as an effective strategy for addressing this particular challenge. An efficient representation of the data may be achieved by utilizing a suitable single-kernel function to map the inherent features and afterward merging them in a space with a high number of dimensions. This approach effectively addresses the differences noted before. The inflexibility of label fitting poses a constraint on the use of multi-kernel late-fusion methods in complex nasopharyngeal carcinoma (NPC) datasets, hence affecting the efficacy of general classifiers in dealing with high-dimensional characteristics. Our approach employs a distinctive framework that incorporates a label-softening technique alongside a multi-kernel-based Radial basis function (RBF) neural network to address this limitation. This innovative methodology aims to increase the disparity between the two cohorts, hence providing a more flexible structure for the allocation of labels. The examination of the NPC-ContraParotid dataset demonstrates the model's robustness and efficacy, indicating its potential as a valuable tool for predicting distant metastases in patients with nasopharyngeal carcinoma (NPC).}




\maketitle

\section{Introduction}\label{introduction}
Asian communities are prone to nasopharyngeal carcinoma (NPC), with a high concentration among Southern Chinese people. Depending on the stage of the disease, different treatment regimens are used. Radiotherapy is used for early-stage diagnoses, and radiotherapy and chemotherapy are combined for more advanced lesions, such as those with nodal involvement or T2--4 disease \cite{al1998chemoradiotherapy, cooper2000improved}. Despite these measures, distant metastasis—a common recurrence pattern in patients with advanced lymph node tumors—often compromises the efficacy of radiation when it is used in combination with other treatments \cite{liu2019treatment, lin2012combined}. The present tumor-nodule-metastasis (TNM) staging system, which serves as the basis for cancer prognostication and directs treatment choices, may contribute to this problem. To correctly anticipate distant metastases, the current method might still need to be improved.

The degree of tumor control in nasopharyngeal cancer has a significant association with the specific radiation dosage applied to the malignant cells. A study conducted on 107 non-small cell lung cancer (NPC) patients showed a significant enhancement in the management of tumor growth when a radiation dosage of 67 Gy was administered \cite{marks1982dose}. In subsequent research including 118 patients diagnosed with nasopharyngeal carcinoma (NPC), it was shown that higher radiation doses and improvements in the technical precision of dose administration were associated with improved tumor control \cite{vikram1985patterns}. The intricacy of radiation planning and administration for nasopharyngeal malignancies stems from the close proximity of the nasopharynx to several vital organs and tissues. This highlights the significance of accurate dosage calibration in treatment protocols. Within this particular setting, radiomics is emerging as a subject increasingly recognized for its utilization of established medical imaging methods, including computed tomography (CT) and magnetic resonance imaging (MRI). The aforementioned pictures function as valuable collections of multi-dimensional data, enabling the identification and analysis of concealed genetic and molecular attributes of malignancies. Consequently, radiomics has the potential to make significant contributions to the development of data-driven prognostic models for cancer. When compared to techniques such as genomes and proteomics, radiomics emerges as a comparatively cost-effective alternative for the prediction of nasopharyngeal carcinoma (NPC) outcomes \cite{zhao2018molecular,zhang2019development,lam2022multi,zhang2023radiomic}.

Although single omics-based approaches have demonstrated potential in predicting distant metastasis \cite{lam2022multi,zhang2022integration} and other diseases \cite{chen2013efficient,chen2016efficient,chen2012support}, their effectiveness is fundamentally constrained by the breadth of information they incorporate. The presence of certain factors has the potential to hinder the acquisition of complete knowledge and restrict the effectiveness of prediction models \cite{zhang2022repeatability,teng2022improving,teng2022building}. Given these circumstances, it becomes apparent that an integrative method incorporating the fusion of multi-omics features presents itself as a more favorable and viable option. The utilization of such a methodology has the potential to improve the clustering of samples and facilitate a more comprehensive comprehension of prognostic and predictive phenotypes \cite{chakraborty2018onco,olivier2019need,hu2021multi}. Recent studies have provided supporting data that highlights the potential of the multi-omics technique in effectively carrying out medical prediction tasks \cite{zhu2022deep,zhou2019latent,dong2023multimodal,lam2022multi,ho2023association,li2022function,zheng2023multi}. For example, the research done by \cite{zhu2022deep} utilized a multilayer perceptron (MLP) to reconstruct individual omics data by employing a hierarchical representation. This approach incorporated shared self-expression coefficients to integrate inter-class and intra-class structural information. The study described by \cite{zhou2019latent} examined the connections between several omics data sets in order to construct a latent representation space that incorporates all available samples.

The investigation of distant metastasis in NPC poses two main obstacles. The initial complexity lies in the link between radiomics, dosiomics, and distant metastases. Although radiomics data contain many variables, not all are pertinent for predicting distant metastases. Moreover, differences may arise across multiple medical pictures due to variances in data collecting time and source for the same patient. The second obstacle emerges from mapping high-dimensional features for the complex original NPC data samples. This highlights the necessity for classification capability when dealing with subsequent works. The objective of conventional multi-kernel learning approaches for classification tasks is to get a transformation matrix that can effectively turn a combination of kernels into a binary label matrix. Nevertheless, the restrictive nature of label fitting and the constrained adaptability of these approaches hinder the efficient exploitation of high-dimensional characteristics by conventional classifiers \cite{chen2011novel,chen2011support,zuo2013effective}.

\begin{figure*}[ht] 
\includegraphics[width=\textwidth]{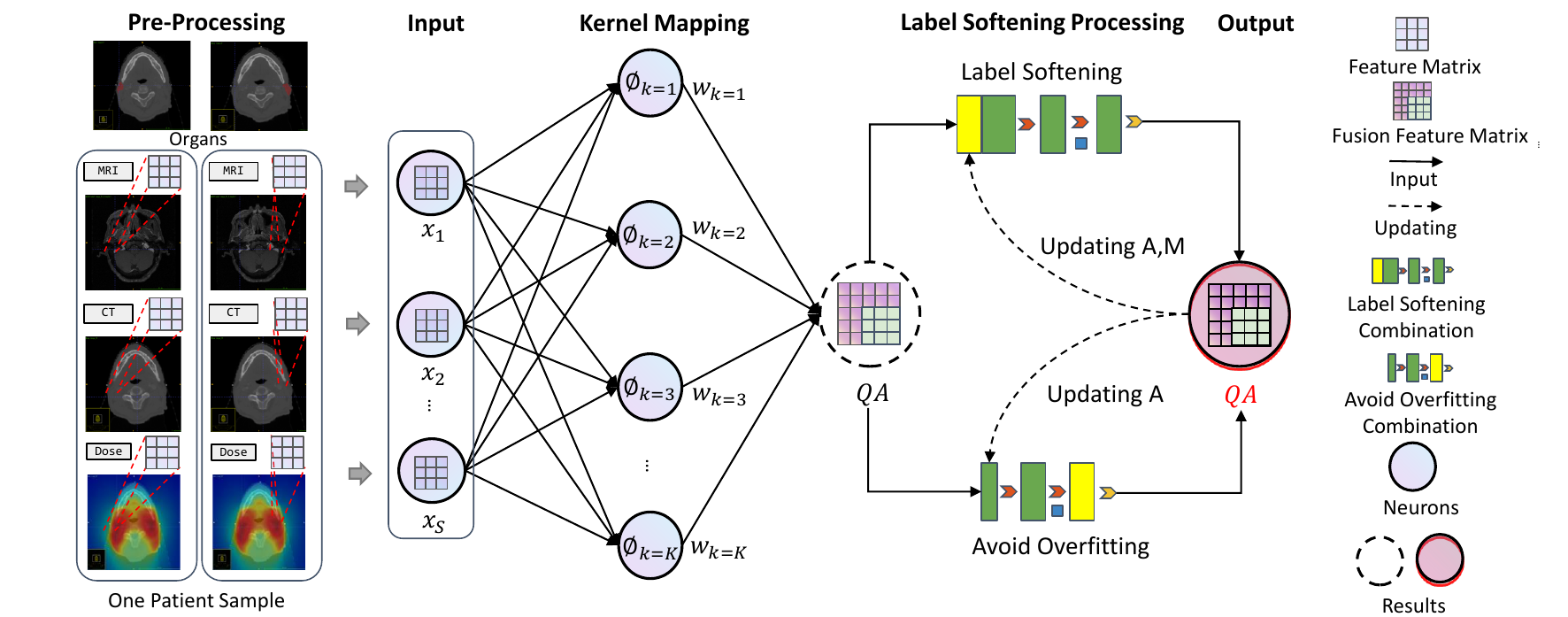} 
\caption{Illustration of the framework. There are five steps, image pre-processing, input, kernel mapping, label softening operation, and result output. We use three modalities of medical images for feature extraction and selection and obtain the results through multi-kernel mapping. Each neuron $k$ contains multiple kernel functions $l \in \{1,L\}$. In classifiers with label softening, the weights are continuously optimized and adjusted through calculation results.} 
\label{framework} 
\end{figure*}

Faced with the difficulties posed by data noise and inconsistent data sources, it is essential to find pragmatic solutions to feature fusion problems. According to \cite{lam2022multi}, their work demonstrates that the utilization of multi-kernel approaches in a late-fusion strategy may successfully mitigate the influence of data inconsistencies, leading to optimized fusion outcomes. This methodology utilizes a multi-kernel technique to create a high-dimensional space that combines many feature spaces. The unique feature mapping capabilities of each subspace are successfully leveraged to facilitate the integration of heterogeneous data obtained from diverse sources. Significantly, each feature is individually mapped using the ideal single kernel function, guaranteeing an accurate and appropriate representation inside the fused space. Furthermore, the study focuses considerable importance on preserving the intrinsic geometric structure of the altered samples \cite{jing2020adaptive}. To achieve this goal, \cite{tian2021large} introduced the margin Fisher analysis method, which leverages label information to maintain the intrinsic geometric and discriminative structures of samples. Certain semi-supervised learning techniques employ an adjacency graph to capture the local structure of samples and maintain label consistency among samples that exhibit similarity \cite{kang2021structured,song2022graph}. The utilization of sample affinity is of utmost importance in order to effectively capture the distribution of samples that belong to the same class, hence assuring their closeness inside the converted space. Figure \ref{framework} a full representation of the processing procedure.

The primary contributions of this study can be summarized as follows:
\begin{itemize}
   \item A novel framework for integrating multi-omics features is proposed. In the pre-processing step, a technique for kernel fusion is employed, which involves the use of matrix-based mixing weights. This approach enables independent learning for each kernel that is involved.

   \item This work presents a meticulously refined and validated prediction model for distant metastasis in nasopharyngeal carcinoma (NPC), utilizing a combination of dose-omics and radiomics. This particular model offers significant insights that have the potential to inform and guide the processes involved in therapeutic decision-making.

   \item This study proposes the use of label softening as a potential solution to the problem of overfitting. The proposed methodology involves transforming the inflexible binary label matrix into a slack variable matrix and constructing a class compactness graph. By solving the regularized minimization issue, it is possible to reach near-optimal margins, hence improving the flexibility and resilience of the model.
\end{itemize}

\section{Related Work}
\subsection{Multi-Kernel Learning}
Kernel functions are highly effective in capturing nonlinear patterns in data. By utilizing the kernel trick, these algorithms implicitly learn nonlinear features in a reproducing kernel Hilbert space (RKHS), even in high-dimensional or infinite-dimensional spaces, while keeping the computation tractable. Multiple kernel learning (MKL) is an approach that combines and selects appropriate kernels for various learning tasks such as classification, clustering, and dimensionality reduction \cite{gonen2011multiple, bucak2013multiple, niazmardi2017multiple}. In recent years, MKL has been extensively researched, resulting in the development of numerous algorithms aimed at enhancing learning efficiency through various optimization techniques \cite{aiolli2015easymkl, alioscha2019svrg}, as well as improving prediction and classification accuracy by exploring different combinations of base kernels \cite{xu2010smooth, varma2009more, cortes2009learning, zhou2016veto}.

Within the domain of multi-omics fusion, the utilization of Multiple Kernel Learning (MKL) offers a resilient framework. A multi-dimensional space, formed by the combination of several kernel functions, integrates various feature spaces. The composite space effectively utilizes the different feature mapping capabilities of each subspace, facilitating the integration of heterogeneous data originating from several sources. The ideal single-kernel function individually maps each feature, resulting in a combined space that provides a more accurate and appropriate representation. \cite{han2013localized} conducted a study on Alzheimer's Disease (AD) identification, wherein they expanded the application of Support Vector Machine (SVM) to the Multi-Kernel Learning (MKL) version. The authors utilized multi-modality data in their research. In this study, kernels were generated utilizing magnetic resonance imaging (MRI), positron emission tomography (PET), and cerebrospinal fluid (CSF) modalities. Similarly, \cite{rakotomamonjy2008simplemkl} employed the "simpleMKL" classifier  to differentiate between individuals with Alzheimer's disease (AD), mild cognitive impairment (MCI), and healthy controls (HCs) using local diffusion tensor imaging (DTI) and magnetic resonance imaging (MRI) modalities. 
However, the initial complex NPC data samples pose some difficulties. More specifically, feature mapping in high-dimensional space does not easily improve the performance of subsequent classification tasks. Typically, in the context of classification problems, multi-Kernel learning focuses on acquiring a transformation matrix that can effectively convert the kernel combination into a matrix of binary labels. The performance of general classifiers is sometimes hindered when used with high-dimensional data due to the inflexibility and limited adaptability of label-fitting methods.

\subsection{Label Softening}
Graph embeddings, an innovative approach drawn from manifold learning, present a promising means to ensure samples bearing identical labels stay proximal post-transformation. This method captures the core data structures effectively while tackling the prevalent issue of overfitting. To combat this, we introduce a 'label blur softening' approach that offers a more flexible binary label matrix. This strategy seeks a new dimensional space where the transformed samples retain the intrinsic geometry of their original counterparts \cite{li2008discriminant}. Yan et al. \cite{yan2006graph} put forth a margin Fisher analysis method, adeptly leveraging label information to uphold both the natural geometric and discriminant structures within the sample. Other methodologies of note include Local Fisher Discriminant Analysis \cite{fan2011local,sugiyama2007dimensionality}, Locality Sensitive Discriminant Analysis \cite{cai2007locality}, and Local Discriminative Embedding \cite{chen2005local}. These approaches share the common objective of ensuring the preservation and usability of geometric and discriminant structures in their newly crafted space.

Semi-supervised learning frequently uses an adjacency graph to encapsulate structure and guarantee label consistency among comparable samples \cite{nie2013adaptive,belkin2006manifold}. However, this technique primarily captures the samples' local structure, thereby curbing its efficiency. To ensure that like-class samples retain proximity in the transformed space, we advocate employing sample affinity to capture the distribution of same-class samples. We introduce the notion of a 'class compactness graph', leveraging label information to connect two samples of the same class via an undirected edge. This strategy ensures the cohesiveness of same-class samples in the transformed space, expanding the gap between disparate categories while providing label-fitting flexibility. This substantially enhances generalization capability and alleviates overfitting concerns. Employing class compactness graphs enables samples—including those distorted by noise—to retain close proximity upon transformation into their label space, thus maintaining a consistent class label. This strategy fosters the creation of an enhanced classifier.

This study presents a label-softening technique within the framework of multi-kernel mapping. After performing high-dimensional mapping, a non-negative label relaxation matrix enables the conversion of a strict binary label matrix into a more flexible slack variable matrix. This transformation effectively mitigates the issue of overfitting, resulting in improved model adaptability. This approach is accomplished by creating class compactness graphs, as seen in Figure \ref{angle}. The solution we propose has two main benefits: firstly, it optimizes the margin between distinct classes to a significant degree; secondly, it offers enhanced adaptability for improved label fitting. By employing a class compactness map, we can preserve the proximity of instances belonging to the same class inside the changed space, successfully mitigating the issue of overfitting. The efficiency of Multi-Kernel Label Softening (MK-FLS) is demonstrated through experimental testing.

\begin{figure}[ht]
  \centering              
    \includegraphics[width=0.3\textwidth]{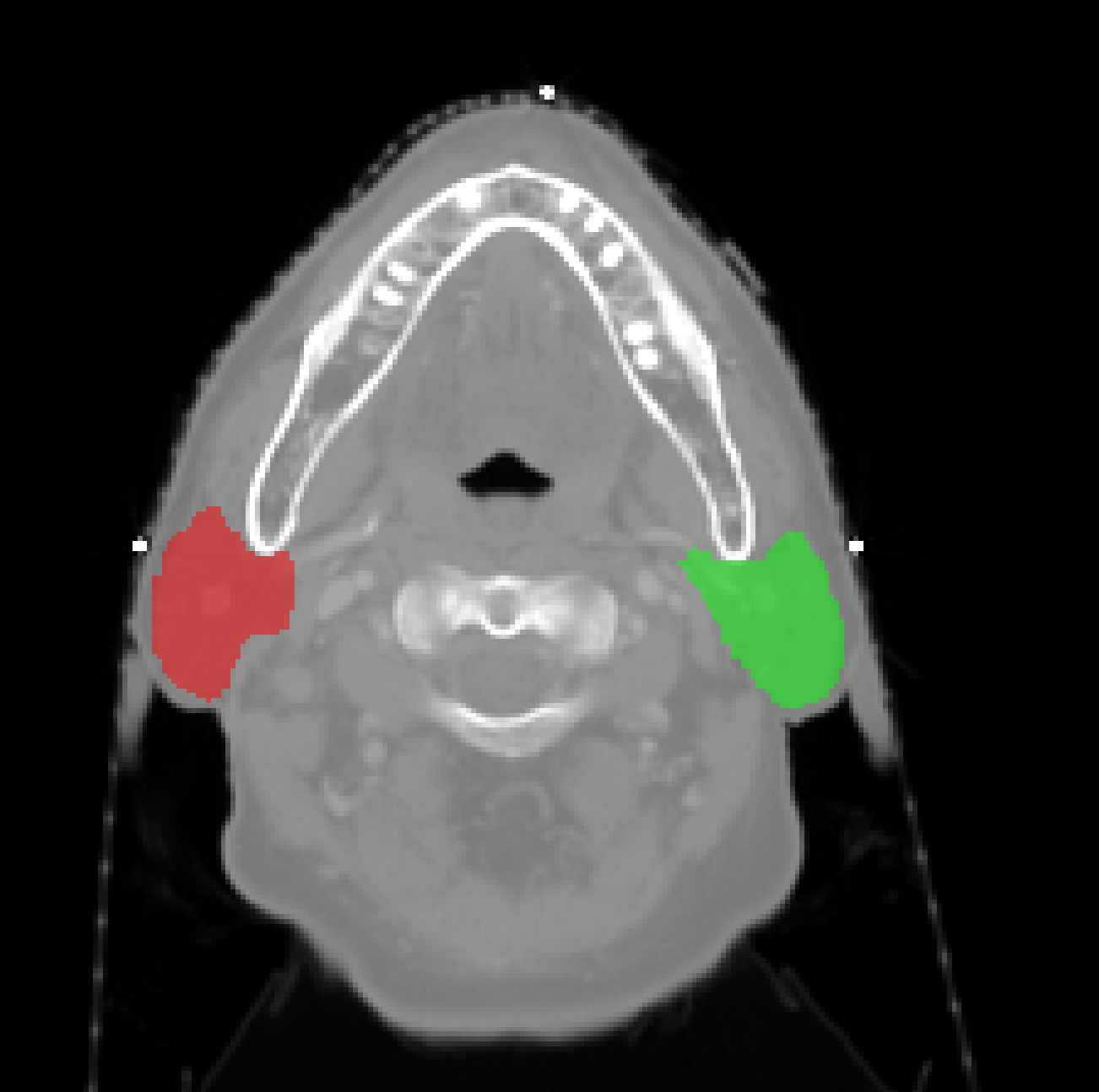}
    \caption{Contralateral parotid glands shown in planning CT images in an example NPC patient.}
    \label{organ}
    \vspace{-3mm}
\end{figure}

\begin{table*}[ht]
\centering
\caption{The statistic of NPC-ContraParotid dataset. The collected images and data are collectively referred to as NPC-ContraParotid, where the dataset includes CT image, MR image, and Dose information. We combined these three different modality permutations to form four sub-datasets. Each sub-dataset is named by the modality type it is composed of.}
\begin{tabular}{cccccrr}
\toprule[1pt]
Organ                         &Datasets&MRI&Dose&CECT&Features&Size\\ 
\hline
\multirow{4}{*}{ContraParotid}&CTD      &   &$\checkmark$&$\checkmark$&10&102\\
                              &MRICT      &$\checkmark$& &$\checkmark$&7&102\\
                              &MRID      &$\checkmark$&$\checkmark$& &11&102 \\
                              &MRICTD      &$\checkmark$&$\checkmark$&$\checkmark$&14&102\\
\toprule[1pt]
\end{tabular}
\label{statistic}
\end{table*}

\section{Materials}
In this part, we will provide a concise overview of the nasopharyngeal carcinoma (NPC) data employed in our studies. This will encompass details on patient characteristics, the technique for image acquisition, the preprocessing procedures applied, as well as the methodologies used for feature extraction and feature selection.

\subsection{Patient information}
The patient data for the planned study will be obtained from our pre-existing database. The data gathered will include planning CT images, patient MR images, and dosage data from the radiation plan. Prior to processing, all patient data will undergo a process of anonymization. In accordance with the specified inclusion criteria, a cohort of 102 eligible patients was enrolled for the purpose of conducting this radiomics investigation. The clinical endpoint that was identified in this study was distant metastases. The confirmation of distant metastatic status for each patient was achieved through fiberoptic endoscopy or histologic/radiographic exams. Patients who experienced the occurrence of distant metastasis subsequent to the completion of the first therapies were designated with a distant metastasis status label of 1. On the contrary, those who maintained a state of being free from sickness until their last follow-up were assigned a value of 0. The duration of the average follow-up period was 36 months. The patient information is presented in Table~\ref{patient}.

\begin{table*}[ht]
\centering
\caption{The patient characteristics of samples.}
\resizebox{\linewidth}{!}{   
\begin{tabular}{ll}
\hline
The inclusion criteria of NPC patients & The exclusion criteria of NPC patients\\ 
\hline
1. Histologically confirmed NPC.                  &1. Palliative treatment was intended.\\
2. Locally advanced Stage I to IV (Cancer stage). &2. Non-primary NPC.\\
3. Availability of pre-treatment MRI and planning&3. Evidence of distant metastasis at diagnosis prior to \\
CT data acquired at the participating hospitals. &   the initial treatment.\\
                                                 &4. Incomplete image data or segment data or clinical data.\\      
                              &5. Neo-adjuvant chemotherapy administrated before \\
                              &chemo-radiotherapy.\\
                              &6. CT images with severe metal artifacts.\\
                              &7. Pretreatment kamofsky performance status (KPS) < 70.\\
                              &8. Existing Comorbidity or pre-existing medical problems.\\
\hline                              
\end{tabular}
}
\label{patient}
\end{table*}

\subsection{Data Processing}
Given the heterogeneity in image capture and reconstruction techniques observed among different medical institutions, it is imperative to engage in pre-processing of images prior to the extraction of radiomic features. This procedural measure enhances consistency, replicability, and soundness in radiomics research. Our study places significant emphasis on four crucial image pre-processing procedures, namely voxel size resampling, VOI re-segmentation, image filtering, and grey-level quantization, in accordance with the standards set out by the Image Biomarker Standardization Initiative \cite{zwanenburg2020image}. These procedures enhance the accuracy and uniformity of our radiomics analysis, consequently improving the overall quality of the results.
\begin{itemize}
   \item Linear interpolation was utilized to resample the CT images, hence modifying the voxel size to $1\times1\times 1 \, mm^3$.
   \item The Hounsfield unit (HU) inside the Volume of Interest (VOI) was restricted to a range of -150 to 180 by the process of VOI re-segmentation.
   \item The Laplacian-of-Gaussian (LoG) filters were implemented, with Gaussian radius parameters of $1 \, mm$, $3 \, mm$, and $6 \, mm$ chosen for the purpose of generating the filtered pictures.
   \item The process of picture grey-level quantization was conducted in order to standardize the signal intensities of the images. In this study, the grey-level intensities of the pictures were discretized using a series of fixed bin count settings. The range of bin counts extended from 50 to 350, with an incremental step of 50.
\end{itemize}

The use of this methodical procedure ensures accuracy and consistency in our research, therefore establishing a strong basis for further analyses.

\subsection{Dataset statistic}
\begin{figure}[ht]
  \centering
    \subfigure[CT]{              
        \includegraphics[width=3.5cm]{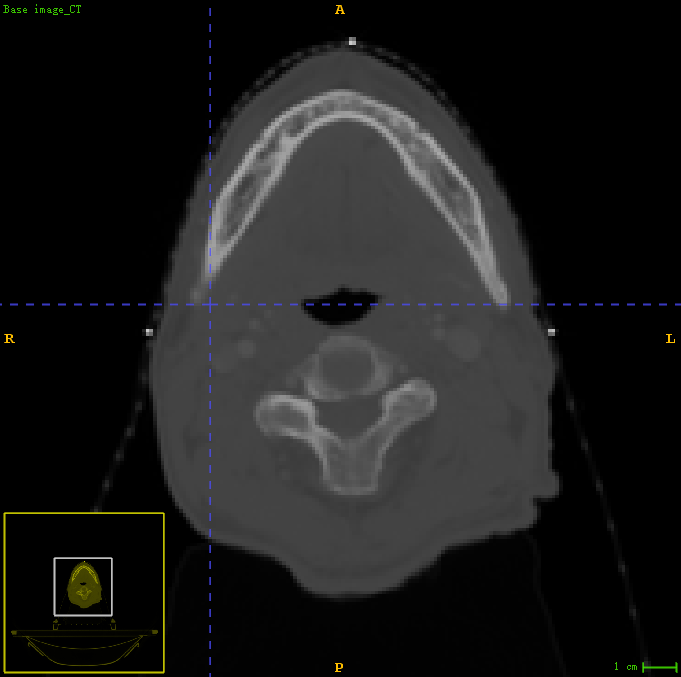}}
    \hspace{5pt}
    \subfigure[MRI(T1C)]{
        \includegraphics[width=3.5cm]{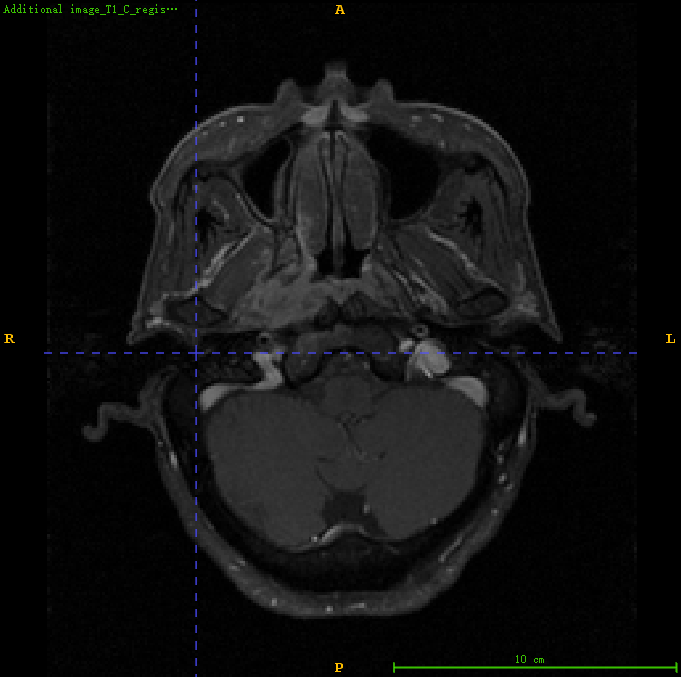}}
    \hspace{5pt}
    \subfigure[MRI(T2)]{
        \includegraphics[width=3.5cm]{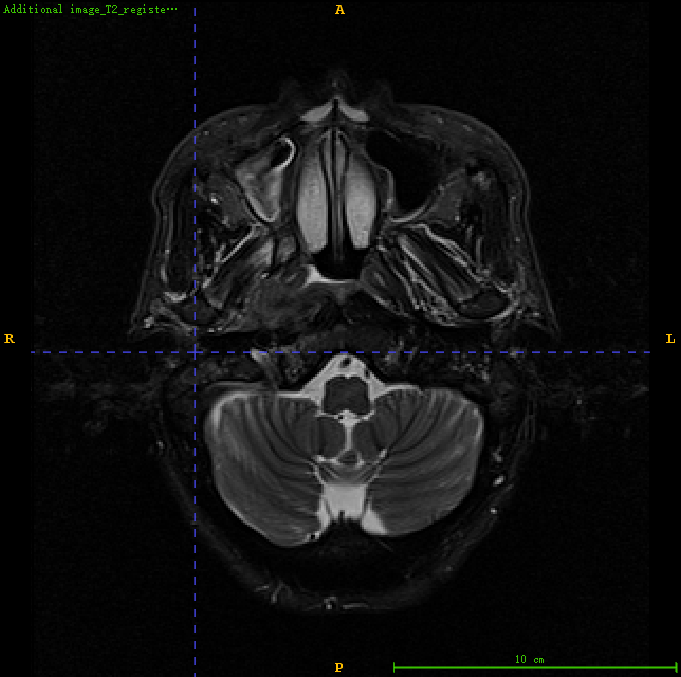}}
    \hspace{5pt}
    \subfigure[Dose]{
        \includegraphics[width=3.5cm,height=3.5cm]{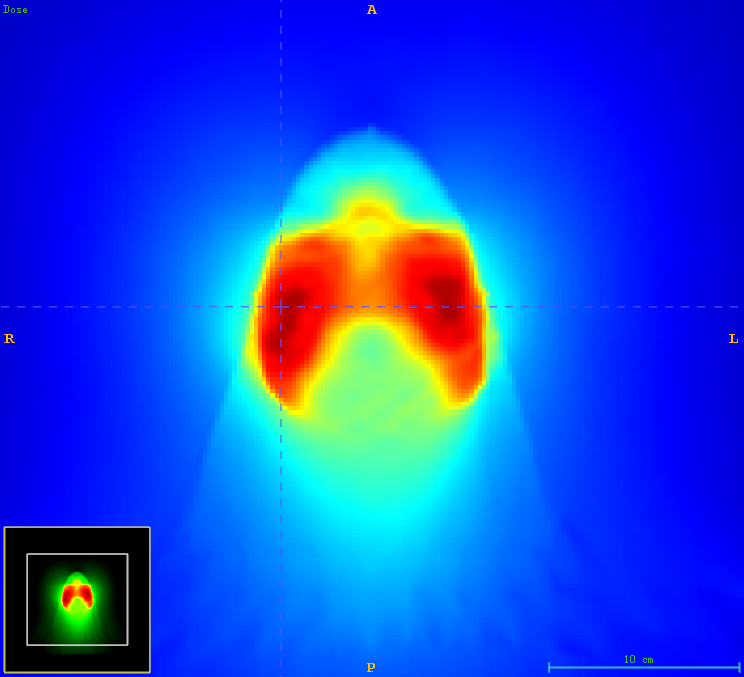}}
    \caption{Samples of CT and MRI images. Although the Dose data is not the medical images, we display the doseomics in the image for understanding}
    \label{sample}
\end{figure}

In order to construct our NPC-ContraParotid dataset, we gathered medical data from a cohort of 102 nasopharyngeal carcinoma (NPC) patients who had radiation therapy in Hong Kong. The dataset consists of three distinct kinds of information pertaining to each patient: Magnetic Resonance pictures (MR pictures), specifically MRI (T1C) and MRI (T2); Computed Tomography (CT) images; and dosage statistics, as seen in Figure \ref{sample}. Patients were categorized as 1 if they had clinically proven distant metastasis and 0 if they did not have any distant metastases. Table \ref{statistic} provides a comprehensive statistical overview of the NPC-to-parotid ratio.

\subsection{Feature Extraction}
In this study, we incorporate the contralateral parotid glands in our radiomic feature calculations, as depicted in Fig.\ref{organ}. We employed SimpleITK v1.2.4 and PyRadiomics v2.2.0—freely available packages—to extract the radiomic features. These packages were integrated seamlessly into our bespoke Python v3.7.3 workflow, thus streamlining our research process and bolstering the reliability of our findings.
\begin{itemize}
  \item The extraction of features was performed on both CT and MR images, employing Laplacian-of-Gaussian (LoG) filters in some cases. The aforementioned attributes may be classified into three overarching categories: shape, first-order statistics, and textural features. In this study, a total of 2130 radiomic characteristics were acquired from both unfiltered and LoG-filtered pictures. The aforementioned characteristics consist of 14 form features, 72 first-order statistics, and 2044 texture features, obtained using predetermined bin count settings. The comprehensive process of feature extraction enhances our capacity for data analysis and broadens the scope of our research discoveries.

  \item Utilizing the dose data, dosiomics characteristics were computed for all organ structures. The conventional utilization of dose-volume histograms (DVHs) is constrained in their capacity to visually represent the spatial dispersion of radiation dosage within affected organs. In accordance with the methodology suggested by Gabrys et al. (2018), we analyzed to ascertain the dosiomics characteristics of DVH curve points for the organs under investigation. These characteristics encompassed the maximum dosage, lowest dose, and mean dose. In order to comprehensively assess the variability of the radiation given, including the gradients of dosage along the three imaging axes (x, y, and z), we acquired the spatial distribution of dosage inside each organ under investigation. The 3D dose distribution of each organ was transformed into a 3D picture, enabling the PyRadiomics program to calculate dosiomics properties similar to radiomics. The aforementioned characteristics include primary dosage statistics, Grey Level Dependence Matrices (GLDM), Grey Level Co-occurrence Matrices (GLCM), Grey Level Run Length Matrices (GLRLM), Grey Level Size Zone Matrices (GLSZM), and Neighboring Grey Tone Difference Matrices (NGTDM). In this study, a total of 1608 dosiomics characteristics were retrieved, resulting in a complete dataset that was utilized for our analysis.
\end{itemize}

\subsection{Feature Selection}
\label{feature selection}
We utilized the Area Under the Curve (AUC) filter \cite{bommert2020benchmark} to evaluate the individual predictive capability of each feature across various omics datasets. The purpose of this filter is to evaluate the precision of classification by examining the performance of each individual feature in predicting class membership. We have developed a prediction rule for the class variable $Y$ based on each feature $X_k$. This rule is represented as $\widehat{Y}=\mathbb{I}_{[c,\infty)}(X_k)$, where $\mathbb{I}$ denotes the indicator function.

The area under the receiver operating characteristic curve (AUC) is a measure of the performance of the classification rule $\widehat{Y}=\mathbb{I}_{[c,\infty)}(X_k)$ in discriminating the target variable. A receiver operating characteristic (ROC) curve with an area under the curve (AUC) value of 1 indicates perfect accuracy, indicating a threshold value of $c$ where the prediction rule is consistently true. On the other hand, when the value is 0, it indicates that the rule inaccurately predicts all labels. This suggests that $X_k$ has the potential to achieve flawless categorization using the rule $\widehat{Y}=\mathbb{I}_{[c,\infty)}(X_k)$. When there is no relationship between the feature and the class variable, the resulting AUC value is 0.5, indicating the lowest possible outcome within this context.

\begin{equation}
    J_{auc}(X_k) = {\arrowvert 0.5 - AUC \arrowvert},
    \label{eq}
\end{equation}

The AUC is performed for the classification rule $\widehat{Y}=\mathbb{I}_{[c,\infty)}(X_k)$, which is utilized as the scoring mechanism for the AUC filter. The characteristics selected from each omics dataset are those that have an area under the receiver operating characteristic curve (AUC) greater than 0.5 and less than or equal to 1. This filter is specifically designed for datasets that consist of two classes.

\begin{figure*}[h] 
\centering 
\includegraphics[width=0.65\textwidth]{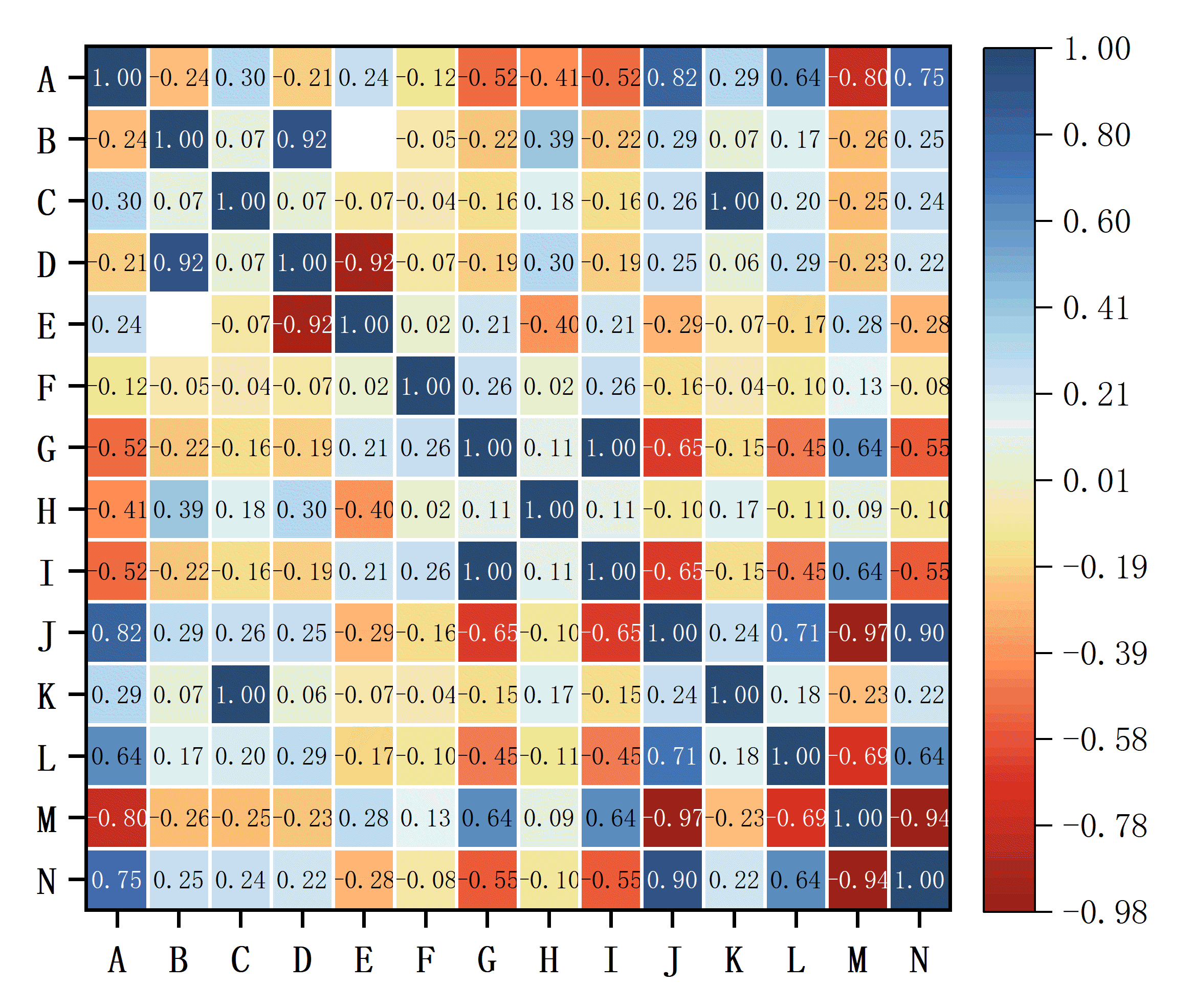} 
\caption{Features correlation of MRICTD in NPC-ContraParotid dataset.The blue block indicates a high correlation coefficient; the closer the correlation coefficient value is to 1, the darker the color. A red block shows a low correlation coefficient; the closer the correlation coefficient value is to -1, the darker the color.}
\label{feature} 
\end{figure*}

A heat map illustrating the relationships between features in the MRICTD dataset was constructed for reference. This heat map is depicted in Figure \ref{feature}. In accordance with the selection process described in Section \ref{feature selection}, a total of 14 separate characteristics were excluded from the analysis. These features were derived from three different types of medical photographs. Table 1 presents a comprehensive list of the specific feature names. The Pearson correlation coefficient (Cohen, 2009) was employed to assess the connection between the various modalities. Based on the visual representation provided, it is evident that while there is a strong connection among variables within the same modality, there were discrepancies between variables from other modalities, leading to diminished correlation coefficient values. In light of this particular attribute of the dataset, it is imperative to utilize model algorithms in order to mitigate the disparities in data distribution across modalities and amplify the impact of various modalities on subsequent tasks.

\begin{table*}[ht]
\caption{The detail selected feature names of MRICTD in NPC-ContraParotid dataset.}
    \centering
\begin{tabular}{lll}
\toprule[1pt]
Features&Image&Original Feature Name\\ 
\hline
\multirow{6}{*}{A-F}& \multirow{6}{*}{Dose} & original$\_$firstorder$\_$90Percentile$\_$1.00$\_$binWidth\\
                    &                       & moment$\_$0$\_$0$\_$3 \\
                    &                       & moment$\_$2$\_$2$\_$0 \\
                    &                       & original$\_$firstorder$\_$10Percentile$\_$1.00$\_$binWidth \\
                    &                       & original$\_$glcm$\_$Imc1$\_$1.00$\_$binWidth \\
                    &                       & original$\_$firstorder$\_$Maximum$\_$1.00$\_$binWidth \\
\multirow{5}{*}{G-K}& \multirow{3}{*}{MRI(T1C)} & wavelet-HHH$\_$glcm$\_$MCC$\_$50$\_$binCount \\ 
                    &                           & log-sigma-3-0-mm-3D$\_$firstorder$\_$MeanAbsoluteDeviation$\_$50$\_$binCount \\
                    &                           & log-sigma-6-0-mm-3D$\_$glszm$\_$SmallAreaLowGrayLevelEmphasis$\_$250$\_$binCount \\
                    & \multirow{2}{*}{MRI(T2)}  & log-sigma-1-0-mm-3D$\_$glcm$\_$Autocorrelation$\_$300$\_$binCount \\
                    &                           & wavelet-LHL$\_$glrlm$\_$LongRunEmphasis$\_$150$\_$binCount \\
\multirow{3}{*}{L-N}& \multirow{3}{*}{CT}   & CT$\_$wavelet-HHL$\_$glszm$\_$ZoneVariance$\_$250$\_$binCount \\
                    &                       & original$\_$firstorder$\_$10Percentile$\_$50$\_$binCount \\
                    &                       & wavelet-HHH$\_$glrlm$\_$HighGrayLevelRunEmphasis$\_$350$\_$binCount \\
\toprule[1pt]
\end{tabular}
\label{feature name}
\end{table*}

\section{Methodology}

\subsection{Multi-Kernel Learning}
The utilization of the multi-kernel learning (MKL) technique offers a robust approach to enhance the performance of mapping optimization. This strategy involves utilizing several or single kernel functions distinguished by varying parameters. Within the extensive range of potential combinations, it has been determined that the linear combination emerges as the prevailing choice. In this study, we consider a set of $L$ distinct kernel functions denoted as $\phi_{l}$, where $l$ denotes the function's index and satisfies the condition $1 \leq l \leq L$. The linear combination of these kernel functions is then expressed as:

\begin{equation}
\bm{\phi} = \sum_{l=1}^{L} \bm{\omega_{l}} \bm{\phi_{l}},
\label{eq1}
\end{equation}
The mixing weight for the $l^{th}$ kernel function is represented by $\bm{\omega}_{l}$ in this expression, whereas the corresponding kernel function is represented by $\bm{\phi}_{l}$. The objective of multi-kernel learning is to acquire the best blending weights for several kernel functions, hence enhancing the efficacy of mapping input data to output.

This study introduces a novel framework for the Radial Basis Function (RBF) neural network architecture \cite{atif2022multi}. The network is composed of an input layer, a nonlinear hidden layer, and a linear output layer, as seen in Figure 1. Suppose Let $X \in \mathbb{R}^{a \times S}$, representing an input dataset of S samples. Each sample $\bm{x}$ in the dataset is a vector in $\mathbb{R}^{a \times 1}$ and is characterized by many characteristics. For each $k$, let $m_k \subset M \in \mathbb{R}^{a \times K}$. Here, $K$ represents the number of neurons in the hidden layer of the radial basis function (RBF) network. In this context, the matrix $M$ is an element of the set of real numbers of $\mathbb{R}^{a \times K}$. It consists of $K$ vectors, denoted as $m_k$, which are individual elements of the set of real numbers of $a \times 1$. These vectors represent the center points of the kernel associated with the $k^{th}$ hidden neuron. Therefore, the expression for the output of each neuron may be stated as follows:
\begin{equation}
\phi_k(\bm{x},\bm{m_k}) = \sum_{l=1}^{L} \omega_{l_k} \phi_{l_k}(\bm{x},\bm{m_k}),
\label{eq2}
\end{equation}
Suppose The symbol $L$ denotes a set of unique kernels in the $k^{th}$ neuron, with $l$ being an element of $L$. In this context, the symbol $\phi_{l_k}$ represents the $l^{th}$ main kernel associated with the $k^{th}$ neuron, while $\bm{\omega}_{l_k}$ denotes the corresponding mixing weight. Two limits are applicable to the vector $\bm{\omega}_{l_k}$: $0 \leq \bm{\omega}_{l_k} \leq 1$, and $\sum_{l}^{L} \bm{\omega}_{l_k}$=1. When paired with these two limitations, the universal set of kernel weights for all multi-kernels guarantees that the participating kernels form a convex combination.
According to the principles of the Radial Basis Function (RBF) technique, the output y for a given input sample x may be mathematically represented as follows:
\begin{equation}
y = \sum^{K}_{k=1} w_k \sum^{L}_{l=1}\omega_{lk}\phi_{lk}(\boldsymbol{x},\boldsymbol{m_k})+b,
\label{eq3}
\end{equation}
Eq.\ref{eq3} may be reformulated as
\begin{equation}
y = \Phi^{\intercal}w
\label{eq4}
\end{equation}
Where $w = [b,w_{11},w_{12}, \ldots ,w_{1K}, \ldots , w_{L1}, w_{L2}, \ldots, w_{LK}]^{\intercal}$, $\Phi=[1, \phi_{11}(\bm{x}, \bm{m_1}), \phi_{12}(\bm{x},\bm{m_2}), \ldots ,\phi_{1K}(\bm{x},\bm{m_K}), \ldots $, $\phi_{L1}(\bm{x},\bm{m_1}),\phi_{L2}(\bm{x},\bm{m_2}), \ldots ,\phi_{LK}(\bm{x},\bm{m_K})]^{\intercal}$. Therefore, for a training set $X=[\bm{x_1},\bm{x_2}, \ldots ,\bm{x_N}]^{\intercal}$, the corresponding output $Y$ can be formulated as
\begin{equation}
\bm{Y}_{mapping} = \bm{QA} 
\label{eq5}
\end{equation}
\begin{equation}
\bm{Q}=
\left[                
  \begin{array}{cccc}   
    1 & \phi_{l1}(\bm{x_1},\bm{m_1}) & ... & \phi_{LK}(\bm{x_1},\bm{m_K})\\  
    1 & \phi_{l1}(\bm{x_2},\bm{m_1}) & ... & \phi_{LK}(\bm{x_2},\bm{m_K})\\
    \vdots & \vdots & \vdots & \vdots\\
    1 & \phi_{l1}(\bm{x_N},\bm{m_1}) & ... & \phi_{LK}(\bm{x_N},\bm{m_K})\\
  \end{array}
\right]
\label{eq6}
\end{equation}
\begin{equation}
\bm{A}=
\left[                
  \begin{array}{cccc}   
    b & w_{l1}(1) & ... & w_{LK}(1)\\  
    b & w_{l1}(2) & ... & w_{LK}(2)\\
    \vdots & \vdots & \vdots & \vdots\\
    b & w_{l1}(N) & ... & w_{LK}(N)\\
  \end{array}
\right]^{\intercal}
\label{eq7}
\end{equation}
The optimization of A can be realized by adopting the following objective, which is premised on the principle of empirical and structural risk minimization:
\begin{equation}
\mathop{min}\limits_{A}{\Vert \bm{QA} - \bm{Y} \Vert}^{2}_{F} + \lambda{\Vert \bm{A} \Vert}^{2}_{F},
\label{eq8}
\end{equation}
Here, $\lambda$ denotes a positive regularization parameter with the condition that $\lambda \ge 0$.

\subsection{Label Softening}
In response to the fact that previous studies have shown that learning discriminative models by overfitting strictly binary label matrices is insufficient, we propose a better method. Our approach softens the label matrix $Y$ using two matrices, $V$ and $Z$. This technique, akin to the strategy used in \cite{fang2017regularized}, aims to expand the margin between classes.
\begin{equation}
   \widehat{Y} = Y+V\bigodot Z,
   \label{eq9}
\end{equation}
In the aforementioned context, the symbol $\bigodot$ represents the Hadamard operator. The matrices $V$ and $Z$ are described in the following manner:
\begin{equation}
\centering
    V_{ij} = 
   \begin{cases}
   +1, & \text{if $y_{ij}$ = 1 }\\
   -1, & \text{if $y_{ij}$ = 0}
   \end{cases}
   \label{eq10}
\end{equation}
\begin{equation}
\centering
   \begin{aligned}
   &Z=
   \left[                
     \begin{array}{ccc}   
       z_{11} & ... & z_{1J}\\  
       \vdots & z_{ij} & \vdots\\  
       z_{I1}& ...& z_{IJ}
     \end{array}
   \right]\\
   &i=1,2,...,I, \\
   &j=1,2,...,J, z_{ij} \geqslant 0, i \ne j.
   \end{aligned}
   \label{eq11}
\end{equation}
After the label has been softened, we proceed to make modifications to the initial multi-Kernel regression model as described in Eq.\ref{eq8}. The resulting modified model is as follows:
\begin{equation}
\begin{aligned}
   &\mathop{min}\limits_{\bm{A},Z}{\Vert \bm{QA} -(\bm{Y}+V \odot Z)\Vert}^{2}_{F} + \lambda{\Vert \bm{A} \Vert}^{2}_{F} \\
   &s.t.  Z \ge 0
   \label{eq12}
\end{aligned}
\end{equation}

\subsection{Overfitting Problem}
While label softening can expand class margins, thereby bolstering classification performance, it also bears the potential risk of overfitting due to its fitting flexibility. Consequently, when pursuing a more discriminative model, it is imperative to curb overfitting. 
To address this, we implement a regularization term method to control data fitting. Initially, we establish an undirected graph that encapsulates the relationships among samples. This can be delineated as follows:

\begin{equation}
\centering
    C_{ij} = 
   \begin{cases}
   e^{\frac{{\Vert \boldsymbol{x_i} - \boldsymbol{x_j}\Vert}^2 }{\sigma}}, \text{if $\boldsymbol{x_i}$ and $\boldsymbol{x_j}$ have the same label}\\
   0
   \end{cases}
   \label{eq13}
\end{equation}
In the notation indicated above, the symbol $\sigma$ denotes the width of the kernel. The variables $x_i$ and $x_j$ represent samples generated using a multi-kernel mapping. These samples are correspondingly associated with the $i$-th row and $j$-th column of the matrix $Q$. Consistent with the aforementioned equation, when two samples belong to the same class, an augmented proximity between them leads to an elevated weight. Conversely, if the samples are from separate classes, the weight defaults to 0. 
Thus, as we transpose the samples into label space, the following objective can be applied to uphold our assumption:

\begin{equation}
\begin{aligned}
\mathop{min}\limits_{\boldsymbol{A},Z}\sum_{ij}{\Vert \boldsymbol{g_j} - \boldsymbol{g_j} \Vert}^{2} C_{ij}= \mathop{min}\limits_{\boldsymbol{A}}tr(\bm{A}^{\intercal}\bm{Q}^{\intercal}L\bm{QA})
   \label{eq14}
\end{aligned}
\end{equation}

\noindent The equation ${\Vert A \Vert}^{2}_{F}=tr(A^{\intercal}A)=tr(AA^{\intercal})$ expresses the relationship between the Frobenius norm ofThe symbol $||A||_F^2$ represents the squared Frobenius norm of matrix A, where the operator $tr(.)$ denotes the trace of a matrix. The notation $g_i=x_iA$ denotes the altered representation of the sample $x_i$. The minimization of the objective function aims to enhance the proximity between vectors $\boldsymbol{g_i}$ and $\boldsymbol{g_j}$ inside the modified space. The symbol $L$ represents the Laplacian matrix, which is obtained by subtracting the matrix $C$ from the diagonal matrix $G$. The computation of the diagonal elements of matrix $G$ is given by the formula $G_{ij} = \sum_{j} \mathbf{C}_{ij}$.

By replacing Eq. \ref{eq14} into Eq. \ref{eq8}, we propose the following goal function for FLS-MK:

\begin{equation}
\begin{aligned}
   &\mathop{min}\limits_{\bm{A},Z} {\Vert \bm{QA}- (\bm{Y}+V\odot Z \Vert}^{2}_{F}+\lambda tr(\bm{A}^{\intercal}\bm{Q}^{\intercal}L\bm{QA})\\
   & s.t.  Z\ge 0
   \label{eq15}
\end{aligned}
\end{equation}
\begin{figure}[ht]
  \centering
    \subfigure[Classifier a]{              
        \includegraphics[width=4cm]{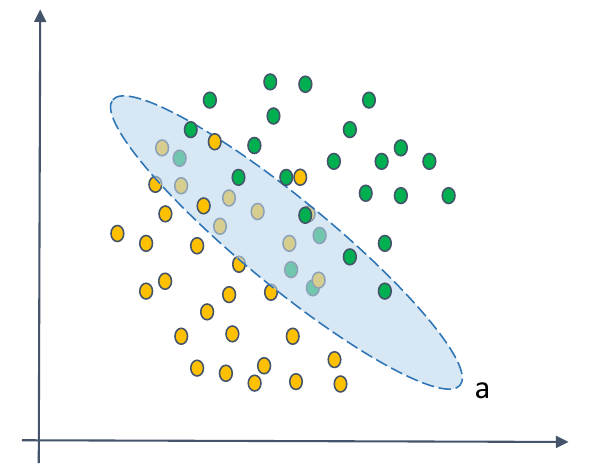}}

    \subfigure[Angle]{
        \includegraphics[width=4cm]{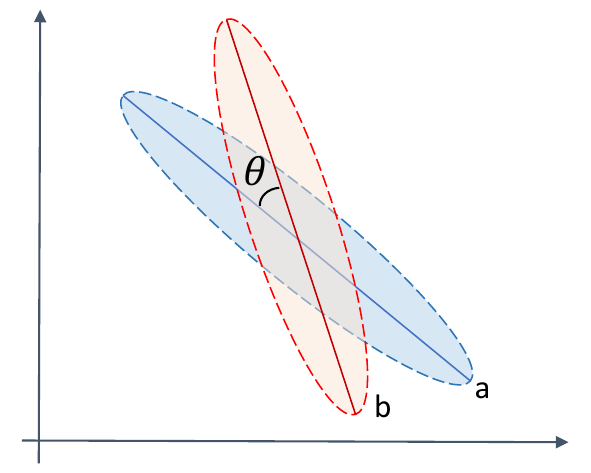}}
    
    \subfigure[Classifier b]{
        \includegraphics[width=4cm]{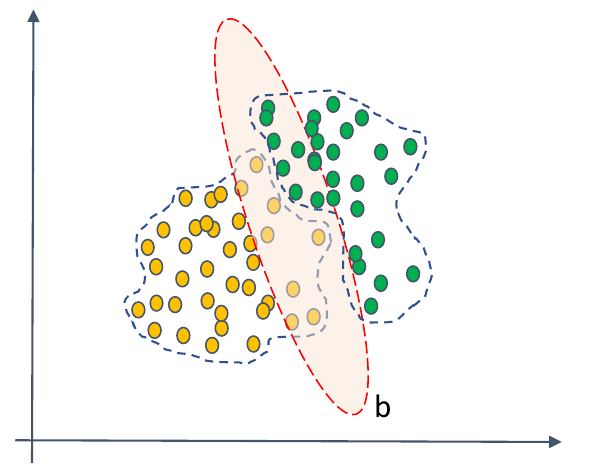}}
    \caption{The provided illustration demonstrates the dynamic alteration in the classifier angle. In pursuit of a broader margin, certain attributes of the training samples may be overlooked during the training process. As displayed in Figure (a), the classifier tends to overfit these anomalous data points due to their significant deviation from regular samples, aiming to distinguish between class 1 and class 2 swiftly. In contrast, Figure (c) employs a class compactness graph to ensure proximity among samples with identical class labels when transformed into label space. This methodology yields a superior classifier 'b'.}
    \label{angle}
\end{figure}

\subsection{Optimization}
In our approach to obtaining the transformation matrix $A$ in Eq. \ref{eq15}, we utilize all training samples without introducing the regularization term ${\Vert A \Vert}^{2}_{F}$ as found in Eq.\ref{eq8}. This method simplifies the process by reducing the need for extensive parameter adjustments.
To illustrate, refer to Fig~\ref{angle}(a). It graphically demonstrates that our refined classifier 'b' is attained by rotating the overfitting classifier 'a' through an angle $theta$. Essentially, the regularization term $tr(\bm{A}^{\intercal}\bm{Q}^{\intercal}L\bm{QA})$ imposes a corrective rotation on the matrix $A$, altering the orientation of the overfit classifier 'a' by an angle $theta$.
This perturbation counteracts the overfitting issue, as depicted in Fig. \ref{angle}(c), making it an effective solution for dealing with such problems in model fitting.

\begin{table*}[ht]
\flushleft
\begin{tabular}{l}
\toprule[1pt]
\emph{\textbf{Algorithm 1:}} Training algorithm of FLS-MK\\ 
\hline
\textbf{Input:}  Training data $ \{\bm{x_i},y_i\}^N_{i=1} $, regularized parameters $\lambda$. \\
\textbf{Output:}  The matrix $A$.\\
\textbf{Procedure:}  \\
  1. Construct the Laplace matrix $L$ according to the training data.\\
  2. Initialize $Z^{(0)}$.\\
  3. $t\gets 0$.\\
  \textbf{Repeat:}\\
  4. Using \\
  \hspace{0.5cm}$A^{t+1}={(Q^{\intercal}Q+\lambda Q^{\intercal}LQ)}^{-1}Q^{\intercal}D^{\intercal}$ to update $A^{(t)}$, where $D^{t}=Y+V\bigodot Z^{t}$.\\
  5. Using \\
  \hspace{0.5cm}$Z=max(V\bigodot R,0)$ to update $Z^{(0)}$.\\
  \textbf{Until} Convergence\\
\toprule[1pt]
\end{tabular}
\label{alg}

\end{table*}

Upon determining the optimal solution $A$, a test sample $x_{test}$ is classified using the following procedure: the transformed result of this sample is given by $y_{test}=x_{test}A$. If $h=\mathop{argmax}\limits_{i} y^{i}_{test} (i=1,..,c)$, then $x_{test}$ is allocated to the $h$-th class. Here, $y^{i}_{test}$ represents the $i$-th element of the vector $y_{test}$. An iterative updating rule is designed to provide a closed-form solution in each iteration and simplify the problem. The algorithm commences by solving for $A$, assuming $Z$ to be fixed. To address this, we apply the following theorem:

\noindent \textbf{Theorem 1:} Given $Z$, and let $\bm{Y}+V\odot Z = D \in \mathcal{R}^{n\times c}$, then the optimal $A$ in Eq.\ref{eq14} can be calculated as
\begin{equation}
\centering
   A = (Q^{\intercal}Q+\lambda Q^{\intercal}\lambda Q)^{-1}Q^{\intercal}D
   \label{eq16}
\end{equation}
\textbf{Proof:} Given an arbitrary $Z$, Eq.\ref{eq14} can be rewritten as
\begin{equation}
    J(A)= \mathop{argmin}\limits_{A}{\Vert QA-D\Vert}^{2}_{F} + \lambda tr(A^{\intercal}Q^{\intercal}LQA)
    \label{eq17}
\end{equation}
In the equation, $D=Y+V\bigodot Z$. The matrix $A$ can be derived by taking the derivative of Eq.\ref{eq16} with respect to $A$ and setting the result equal to zero, as demonstrated below:
\begin{equation}
\begin{split}
    &\frac{\partial J}{\partial A}=2Q^{\intercal}QA-2Q^{\intercal}D+2\lambda Q^{\intercal}LQA=0\\
    &\Rightarrow A=(Q^{\intercal}Q+\lambda Q^{\intercal}LQ)^{-1}Q^{\intercal}(Y+V\bigodot Z)
\label{eq18}
\end{split}
\end{equation}

\noindent \textbf{Theorem 2:} Given fixing $A$, let $QA-Y=R$. According to Eq.\ref{eq13}, the optimal solution of $Z$ is
\begin{equation}
    Z=max(V \bigodot R,0)
\label{eq19}
\end{equation}
\textbf{Proof:} Since $A^{\intercal}Q^{\intercal}LQA$ is uncorrelated with $Z$, Eq.\ref{eq15} can be rewritten as 
\begin{equation}
\begin{split}
        &\mathop{ming}\limits_{Z}{\Vert R-V\bigodot Z \Vert}^{2}_{F}\\
        &s.t. Z \ge 0.
\end{split}
\label{eq20}
\end{equation}

Therefore, $Z$ is finally obtained by Eq.\ref{eq19}.

\noindent \textbf{Theorem 3:} Eq.\ref{eq15} is convex with at least one minimum for any positive $\lambda$. Provided the covariance matrix $Q^{\intercal}Q$ is nonsingular, Eq.\ref{eq15} possesses a unique minimum.

\noindent \textbf{Proof:} Since both $Q^{\intercal}Q$ and $Q^{\intercal}LQ$ are positive semidefinite, Eq.\ref{eq15} becomes a convex function in terms of variable $A$. Given that the matrix $V$ does not include any zero entries, it can be shown that Equation \ref{eq15} becomes a strictly convex function in relation to the variable $Z$. Moreover, the property of having a non-zero determinant for the matrix $Q^{\intercal}Q$ provides confirmation that Eq.\ref{eq15} exhibits tight convexity with respect to both variables $A$ and $Z$. Hence, it can be observed that Eq.\ref{eq15} attains a solitary minimum point, as confirmed by Bertsekas in 2003 (Proposition 2.1.2).

\subsection{Algorithm}
The method of our proposed model is outlined in Algorithm 1. The temporal complexity of the MK-FLS algorithm is mainly determined by the operations required for updating the variables $A$ and $Z$. The temporal complexity of updating matrix $A$ is asymptotically bounded by $O(N^{3})$ due to the requirement of performing matrix inversion. On the other hand, the temporal complexity of updating $Z$ is reduced to a more reasonable $O(N)$. Hence, the estimated asymptotic time complexity for the MK-FLS model is $O(N^{3})$. The MK-FLS model's computational efficiency is improved, increasing its application and performance when dealing with more extensive datasets.

\begin{table*}[h]
\caption{The ACC(\%) of NPC-ContraParotid.}
   \centering
\begin{tabular}{crrrrr}
\toprule[1pt]
Modalities & Ridge Regression & RBFNN-MK &FSVMs  &FSVMs-CIP &FLS-MK\\ 
\hline
CTD         &91.1765          &79.4118   &86.1354&82.3529       &91.6670\\
MRICT         &94.1176          &76.4706   &82.3529&88.2358       &89.5449\\
MRID         &91.1765          &55.8824   &88.2353&85.2941       &95.8940\\
MRICTD         &82.3529          &58.8235   &77.6228&79.4118       &96.0723\\
\toprule[1pt]
\end{tabular}
\label{acc}
\end{table*}

\begin{table*}[ht]
\caption{The AUC of NPC-ContraParotid.}
    \centering
\begin{tabular}{crrrrr}
\toprule[1pt]
Modalities & Ridge Regression & RBFNN-MK &FSVMs  &FSVMs-CIP &FLS-MK\\ 
\hline
CTD         &0.8635           &0.9086    &0.6428 &0.6471    &0.9404\\
MRICT       &0.8824           &0.9123    &0.6644 &0.7751    &0.8812\\
MRID        &0.8304           &0.7612    &0.7647 &0.7059    &0.9471\\
MRICTD      &0.6782           &0.7474    &0.5736 &0.5882    &0.9497\\
\toprule[1pt]
\end{tabular}
\label{auc}
\end{table*}

\section{Experiment}
We curated a collection of four multi-omics datasets, with each dataset including a total of 102 samples (see Table \ref{statistic}). The establishment of a comprehensive collection of features was achieved by the extraction of radiomic features from several sources, namely the planning dosage map, pre-treatment contrast-enhanced CT scans, and T1-weighted contrast-enhanced MR images. These sources were part of the ContrasParid dataset. To showcase the classification capabilities of the proposed model, we selected four models for evaluation. These include two improved fuzzy classification models based on Support Vector Machines (FSVMs \cite{lin2002fuzzy} and FSVMs-CIP \cite{batuwita2010fsvm}), RBFNN-MK \cite{atif2022multi} to demonstrate the impact of label softening, and Ridge Regression \cite{mcdonald2009ridge} as the baseline model for regression analysis.

\subsection{Settings}
In both the training and testing steps, we implemented a K-fold cross-validation technique with a value of k equal to 3. The performance of the proposed model was evaluated using Accuracy (ACC) and the Area Under the Receiver Operating Characteristic Curve (AUC) as the principal criteria.
Accuracy was computed as the ratio of correctly classified samples to the total number of samples, providing a direct measure of our model's proficiency. In contrast, AUC served as a commonly employed metric for gauging the effectiveness of binary classification models. Specifically, it indicates the probability that a positively labeled example will be ranked above a negatively labeled one, offering insights into the classification power of model.

\subsection{Experiment Analysis}
\begin{figure}[ht]
	\centering
	\subfigure{
		\begin{minipage}{0.95\linewidth}
			\includegraphics[width=\textwidth]{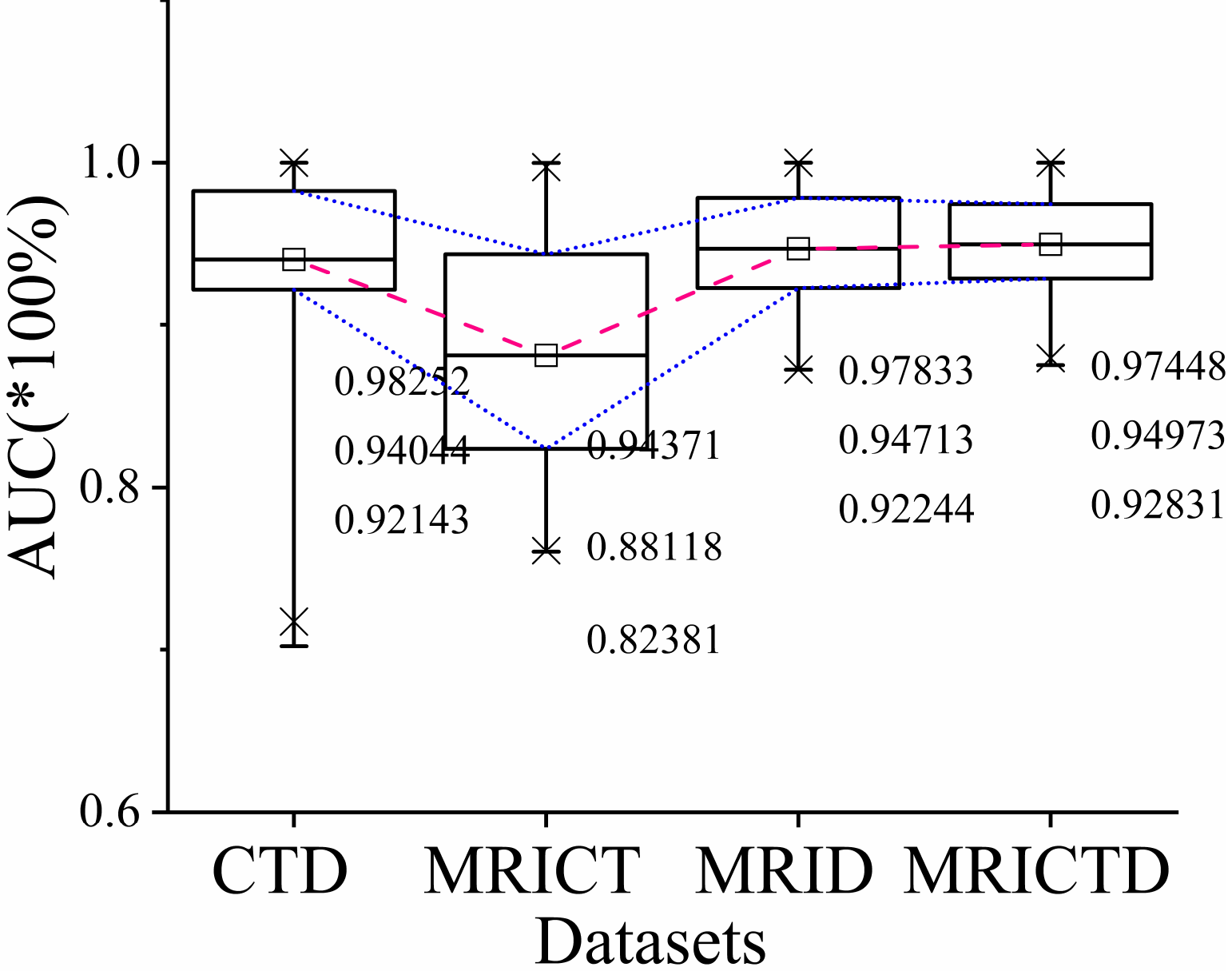}		
		\end{minipage}
	}
	\subfigure{
		\begin{minipage}{0.95\linewidth}
			\includegraphics[width=\textwidth]{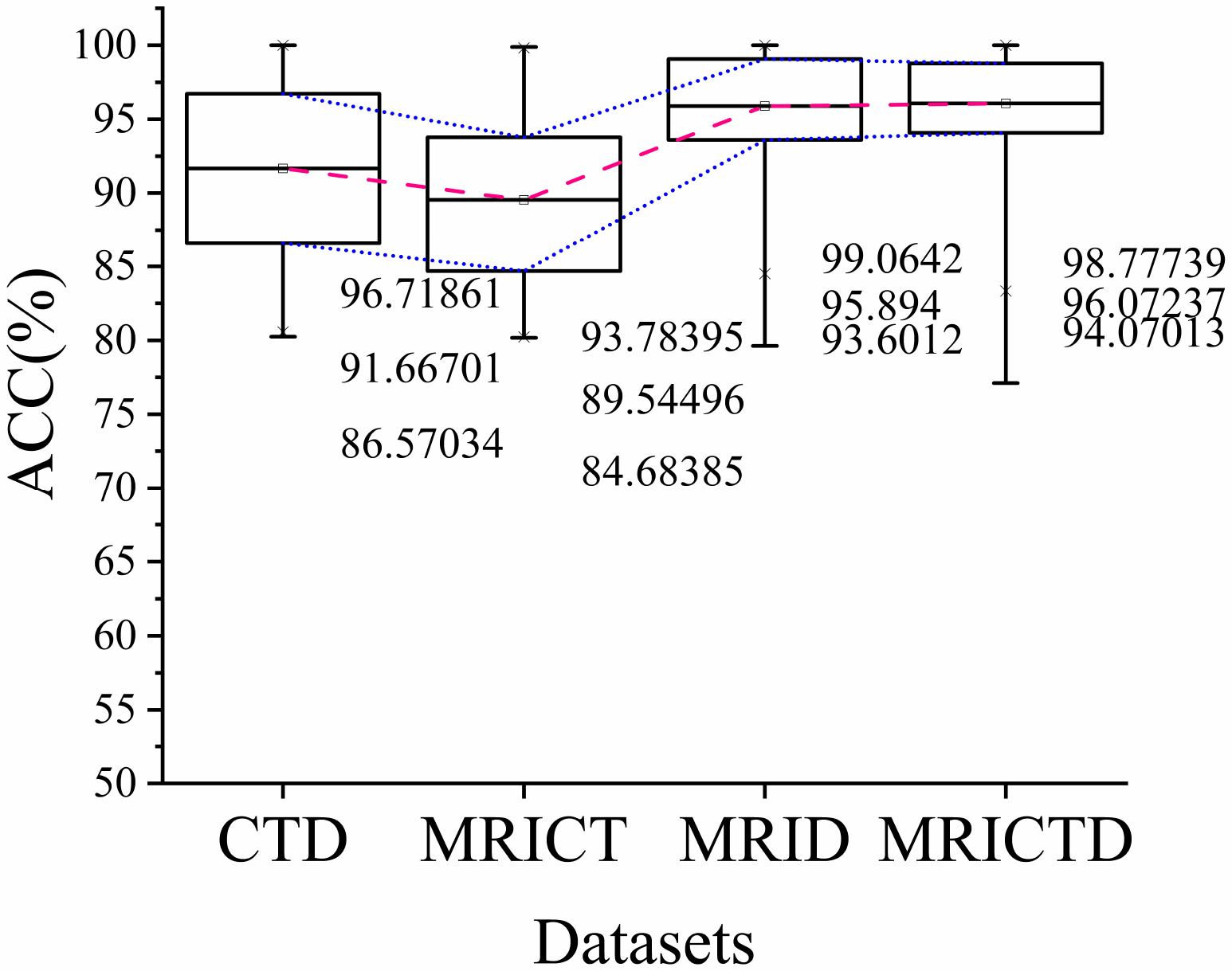}
		\end{minipage}
	}
\caption{The results of FLS-MK on NPC-ContraPariod.}
\label{boxplot}
\end{figure}

\begin{figure}[ht]
	\centering
	\subfigure{
		\begin{minipage}{0.95\linewidth}
			\includegraphics[width=\textwidth]{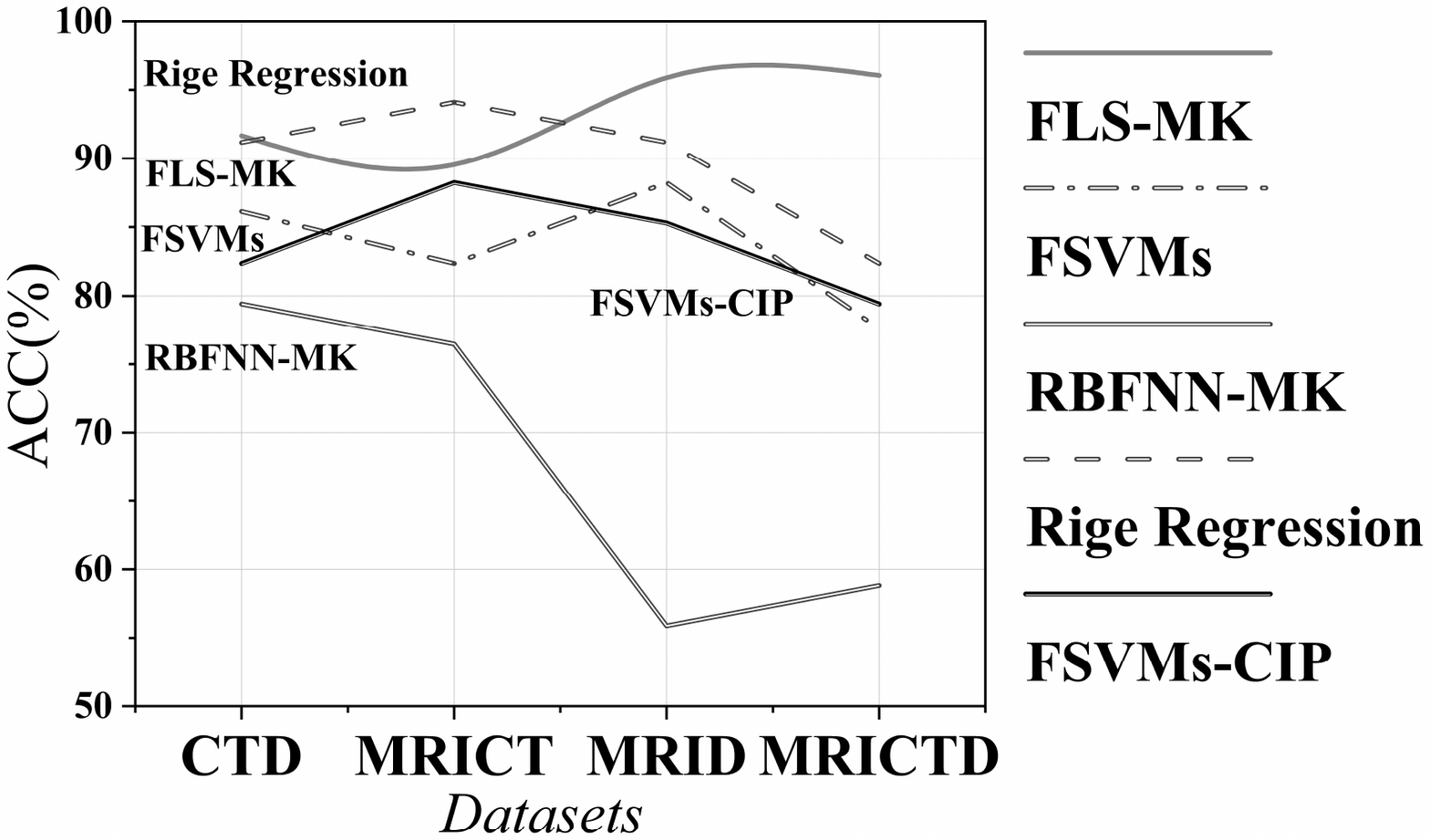}		
		\end{minipage}
	}
	\subfigure{
		\begin{minipage}{0.95\linewidth}
			\includegraphics[width=\textwidth]{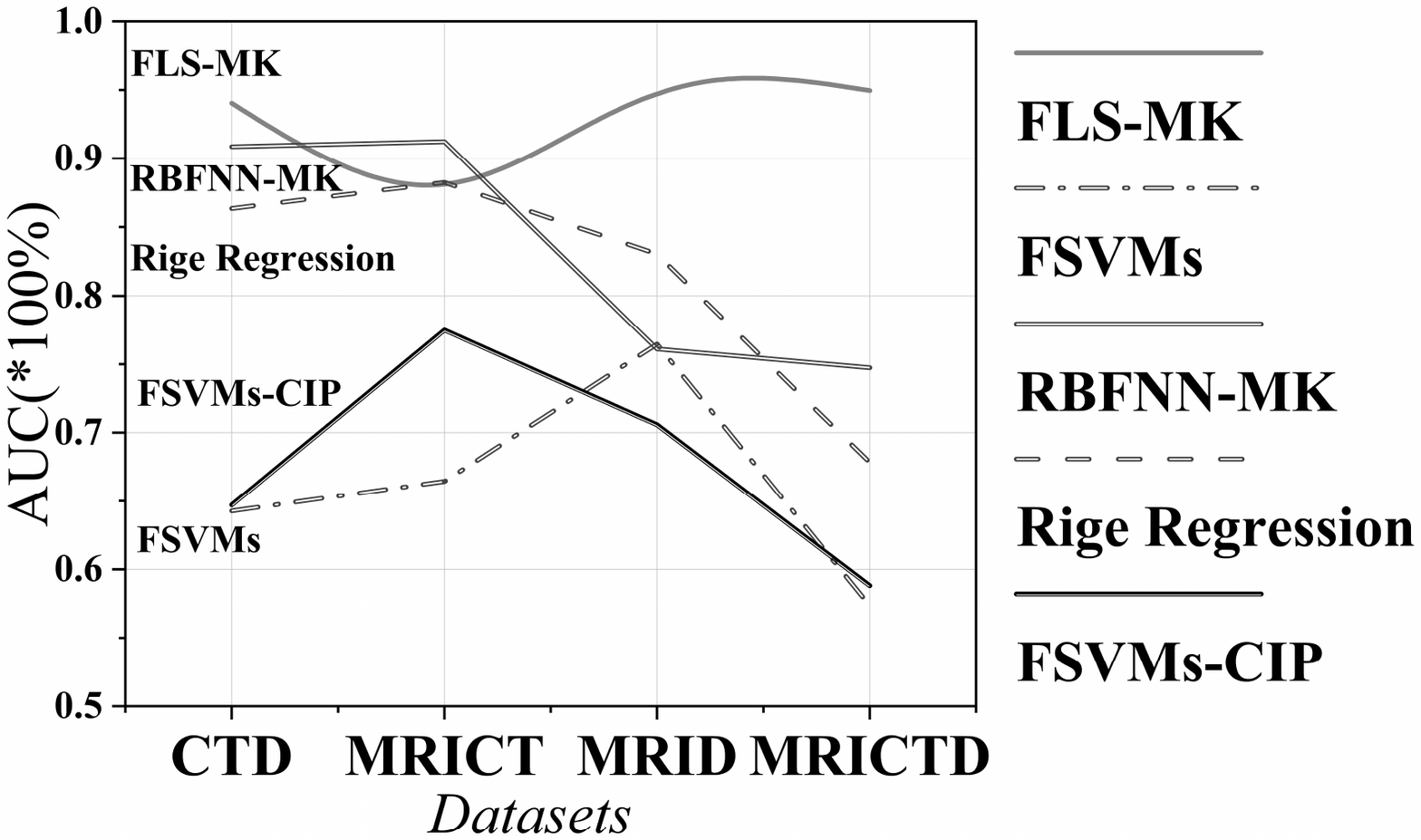}
		\end{minipage}
	}
\caption{Experimental results compared on different omics datasets}
\label{result1}
\vspace{-3mm}
\end{figure}

The FLS-MK model has superior Accuracy (ACC) performance compared to other models when evaluated on the CTD, MRID, and MRICTD datasets, as presented in Table~\ref{acc}. Despite the ACC score of the FLS-MK model on the MRICT dataset not surpassing that of ridge regression, an examination of the feature count in the dataset indicates that FLS-MK exhibits higher effectiveness in incorporating data with numerous quantitative characteristics, as seen in Figure \ref{boxplot}. The Ridge regression, FSVMs, and FSVMs-CIP models achieve accuracy (ACC) values beyond 90\% and 80\% on the CTD, MRICT, and MRID datasets, respectively. These datasets consist of two modalities. However, the accuracy of the models significantly decreases when applied to the MRICTD dataset, which consists of more than two modalities. On the other hand, FLS-MK demonstrates the ability to maintain a high score, suggesting that the classification model, which becomes more flexible after high-latitude mapping, is better suited for multi-modal datasets (as seen in Figure \ref{result1}). In contrast, the RBFNN-MK model exhibits the lowest score compared to other models. This implies that relying solely on high-latitude mapping does not substantially improve the performance of classification tasks.

According to the AUC findings shown in Table~\ref{auc}, it can be observed that the Ridge Regression, FSVMs, and FSVMs-CIP models exhibit a notable impact on datasets that incorporate two modalities. However, the area under the receiver operating characteristic curve (AUC) scores exhibit a significant drop when confronted with datasets that consist of more than two modalities. Significantly, the FSL-MK model has the ability to consistently maintain stable scores across all four segmented datasets, with a notable peak score of 0.9497. The evaluation of AUC and ACC metrics for RBFNN-MK on the four datasets confirms the benefits of combining label softening classification with high-latitude mapping.

\begin{figure}[ht]
\vspace{-2mm}
	\centering
	\subfigure{
		\begin{minipage}{0.95\linewidth}
			\includegraphics[width=\textwidth]{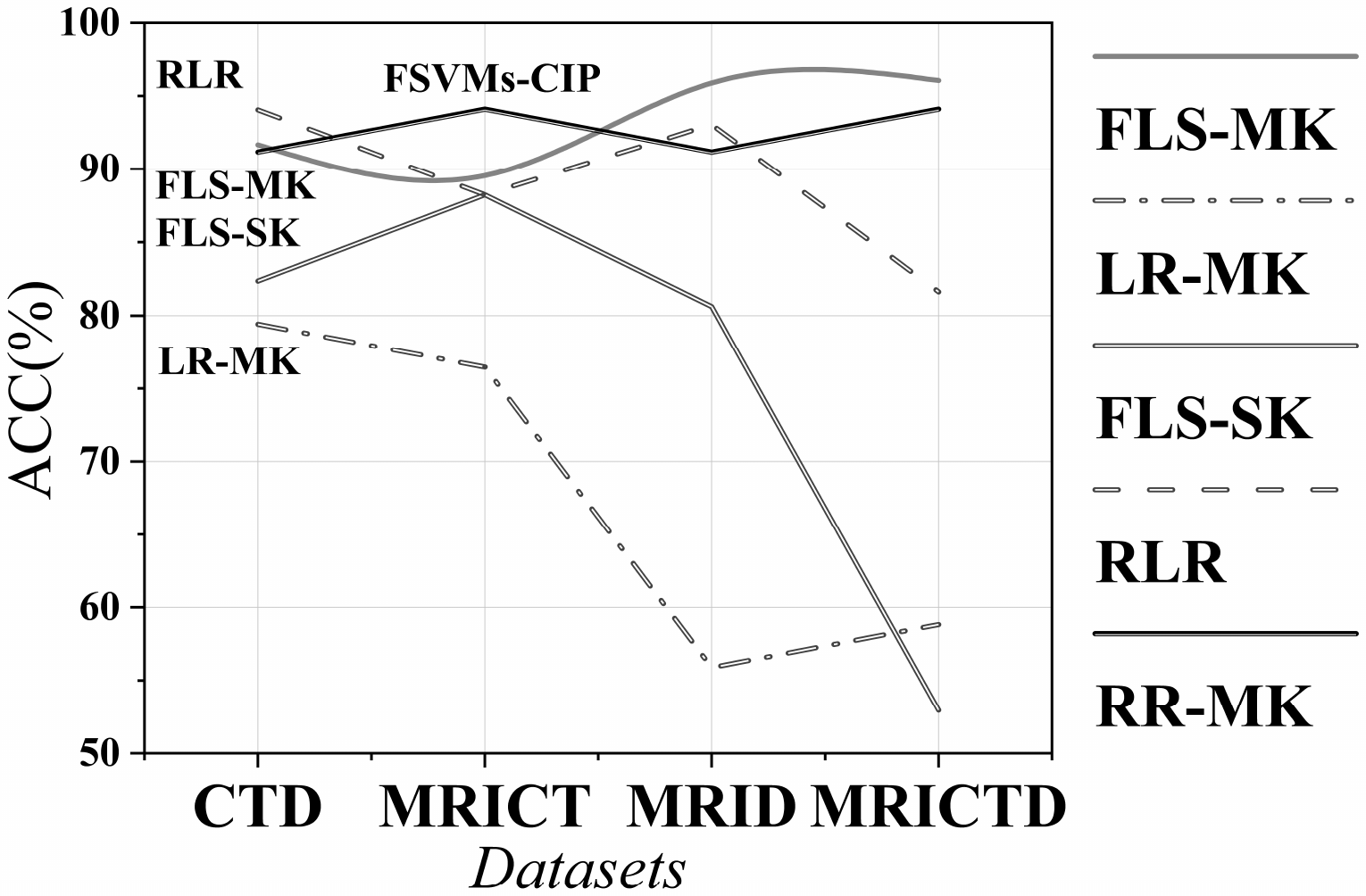}		
		\end{minipage}
	}
	\subfigure{
		\begin{minipage}{0.95\linewidth}
			\includegraphics[width=\textwidth]{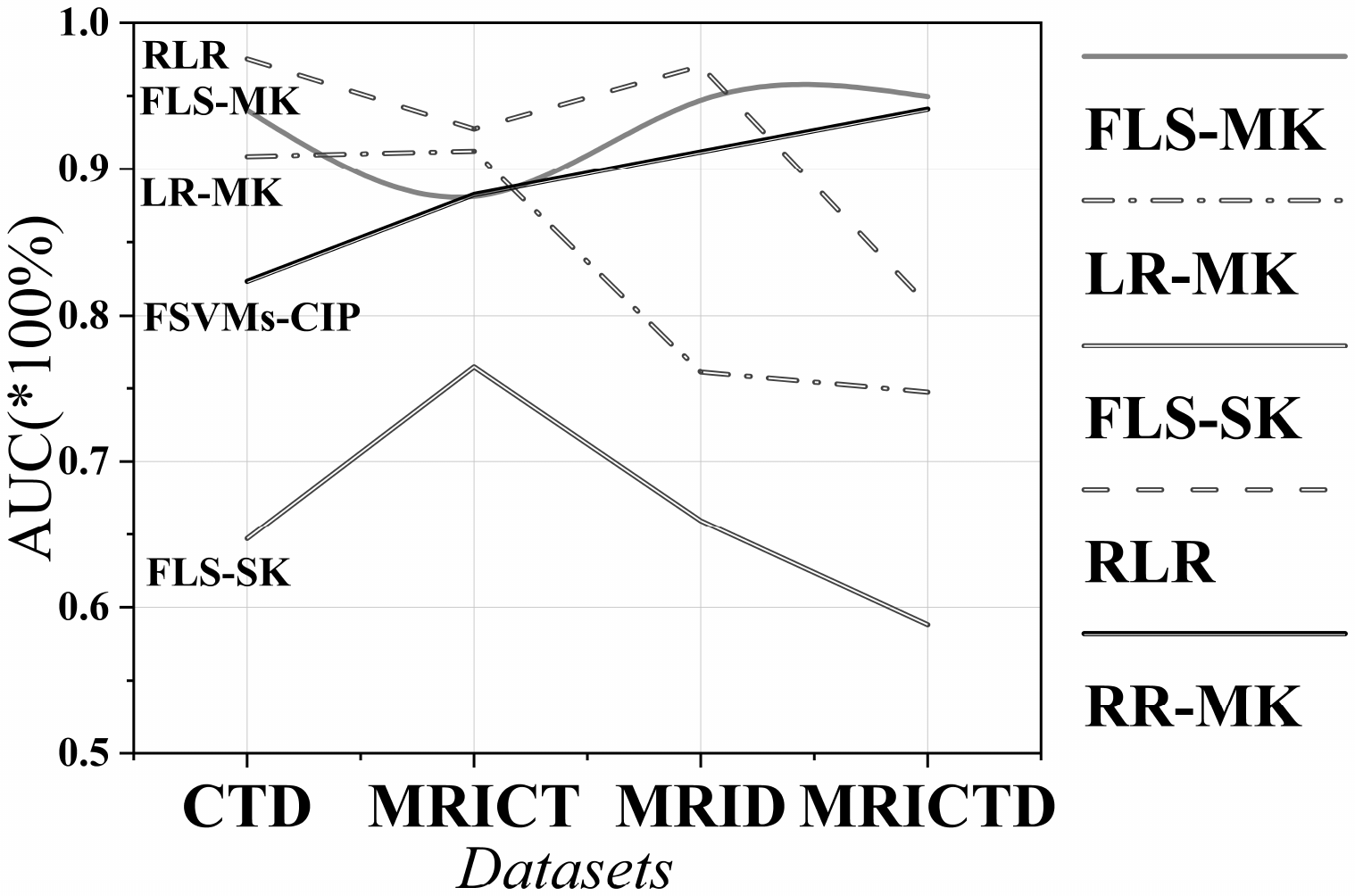}
		\end{minipage}
	}
\caption{Ablation experimental results compared on different omics datasets}
\label{result2}
\vspace{-3mm}
\end{figure}

\subsection{Ablation Studys}
Table~\ref{acc2} and Table~\ref{auc2} show that the multi-kernel-based linear regression classifier (LR-MK) and ridge regression classifier (RR-MK) are used as the benchmark for comparison. Compared with LR-MK, FLS-MK has better classification performance and stability. Compared with the RR-MK method, due to the omics characteristics, the ACC and AUC scores of FLS-MK on the MRICT dataset are lower than those of the RR-MK model. However, the average classification performance on the rest of the multi-omics data is superior. Therefore, the improved  label softening method based on the ridge regression algorithm is valuable for downstream classification tasks.

Comparing the  label softening method based on single-kernel mapping (FLS-SK) and the regularization label relaxation (RLR) classifier \cite{fang2017regularized} without multi-kernel mapping, the method with multi-kernel mapping has superior results in the fusion of two or more. (Here, we select the cosine with the best score in the kernel function as the single-core verification of the ablation experiment.) Although the single-kernel mapping method can also obtain good accuracy results, its classification and multi-omics fusion performance could be better. The RLR classifier without multi-kernel mapping shows the best AUC score on the two omics datasets, up to 0.9754. Nevertheless, it significantly performs poorly for the fusion of more than two omics (shown as Fig.\ref{result2}). Therefore, the FLS method based on multi-core mapping has a more stable classification ability for multi-omics fusion.

\begin{table*}[ht]
\centering
\caption{The ACC(\%) of NPC-ContraParotid.}
\begin{tabular}{crrrrr}
\toprule[1pt]
Modalities & RLR & FLS-SK &LR-MK  &RR-MK &FLS-MK\\ 
\hline
CTD         &94.0582          &82.3529   &79.4118&91.1765       &91.6670\\
MRICT       &88.2353          &88.2353   &76.4706&94.1176       &89.5449\\
MRID        &93.0588          &80.6225   &55.8824&91.1765       &95.8940\\
MRICTD      &81.6228          &52.9412   &58.8235&94.1176       &96.0723\\
\toprule[1pt]
\end{tabular}
\label{acc2}
\end{table*}

\begin{table*}[ht]
\centering
\caption{The AUC of NPC-ContraParotid.}
\begin{tabular}{crrrrr}
\toprule[1pt]
Modalities & RLR & FLS-SK &LR-MK  &RR-MK &FLS-MK\\ 
\hline
CTD         &0.9754           &0.6471    &0.9086 &0.8235    &0.9404\\
MRICT       &0.9276           &0.7647    &0.9123 &0.8824    &0.8812\\
MRID        &0.9706           &0.6594    &0.7612 &0.9118    &0.9471\\
MRICTD      &0.8075           &0.5882    &0.7474 &0.9412    &0.9497\\
\toprule[1pt]
\end{tabular}

\label{auc2}

\end{table*}

\subsection{Data distribution Analysis}
\begin{figure}[ht]

	\subfigure[Before]{
		\begin{minipage}[b]{\linewidth}
            \centering
			\includegraphics[width=0.7\textwidth]{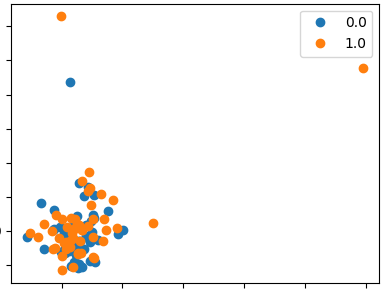}
		\end{minipage}
	}
	\subfigure[After]{
		\begin{minipage}[b]{\linewidth}
            \centering
			\includegraphics[width=0.7\textwidth]{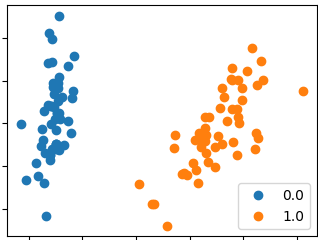}
		\end{minipage}
	}
\caption{The sample distribution of MRICTD dataset in NPC-ContraPariod.}
\label{distribution}
\end{figure}

Figure~\ref{distribution} depicts the initial distribution of the NPC patient dataset alongside the distribution subsequent to the training process. The MRICTD dataset's visualization was achieved using the Principal Component Analysis (PCA) method for dimensionality reduction \cite{kurita2019principal}. As displayed in Fig.~\ref{distribution}(a), segregating the initial dataset into two classes presents a challenge. Nevertheless, as illustrated in Fig~\ref{distribution}(b), superior classification outcomes are obtained following the training process.

\section{Discussion}
This study introduces a label-softening approach incorporated within a multi-kernel mapping architecture. The proposed methodology involves the utilization of a non-negative relaxation matrix to modify the labels after undergoing high-dimensional mapping. This process effectively converts the inflexible binary label matrix into a more flexible variable matrix, hence addressing the issue of overfitting. In our methodology, we incorporate a class compactness map to guarantee that instances belonging to the same class are in close proximity inside the transformed space. The approach developed in this study demonstrates enhanced resilience and is less susceptible to fluctuations in the distribution of information across various modalities. The performance of our method in classification tasks, as evidenced by the AUC score reported in Table~\ref{auc2}, is commendable compared to other techniques such as Regularized Logistic Regression (RLR). Specifically, the RLR algorithm has difficulties in effectively integrating multi-omics data, resulting in a significant decline in its accuracy. Conversely, our suggested technique demonstrates a constant level of performance. In the preliminary assessments, we employed four distinct models: two refined iterations of fuzzy classification models that rely on Support Vector Machines (FSVMs and FSVMs-CIP), Radial Basis Function Neural Networks upgraded with multi-kernel mapping (RBFNN-MK), and Ridge Regression. The present work introduces the Learning System with Multi-kernel mapping (FLS-MK) model, which exhibits exceptional Accuracy (ACC) performance on the CTD, MRID, and MRICTD datasets, as seen in Table~\ref{acc}. The approach employed in this study demonstrates a high level of robustness when applied to combined datasets encompassing multiple omics. This robustness is evidenced by the reduced variability observed in the final results. The utilization of multi-kernel mapping enables the creation of a high-dimensional depiction of the distribution of samples, hence reducing discrepancies in sample distributions across several modalities. Our proposed approach also successfully mitigates the problem of overfitting that may arise due to label softening. In order to enhance the reliability and effectiveness of our approach, we performed ablation experiments. The outcomes of these testing are illustrated in Figure ~\ref{result2}. These tests provide more evidence to support the validity of our strategy.

\section{Conclusions}
This study presents a novel approach, the Multi-Kernel model-based Label Softening (FLS-MK) technique. The proposed methodology utilizes kernel mapping techniques to transform the input data into a higher-dimensional space. Additionally, a positive matrix is employed to regulate the binary label matrix. Consequently, this process enables improved label fitting and increases the margins between different classes. The effectiveness of our methodology has been substantiated by several experiments conducted on multi-omics datasets, as well as by comparing its performance to that of other models and a baseline model. The efficacy of our suggested methodology on the NPC dataset has been successfully shown. This study introduces a promising approach to integrating intricate medical data in the context of illness prediction, enhancing the existing body of knowledge in this domain.

\newpage
\section{Acknowledgment}
We would like to note that a condensed version of this research was previously presented at the 2023 24th International Conference on Digital Signal Processing. However, the current manuscript offers a more comprehensive and in-depth analysis of the study, providing additional insights and findings not included in the initial presentation.

We acknowledge the following participants (from the Department of Clinical Oncology, Queen Elizabeth Hospital, Hong Kong, Hong Kong SAR, China) who contributed to our work by offering administrative and material support for clinical data and imaging data collection: Francis Kar Ho LEE, Celia Wai Yi YIP, and Kwok Hung AU.
This work was supported in part by the Project of RISA (P0043001) of The Hong Kong Polytechnic University, Shenzhen-Hong Kong-Macau S\&T Program (Category C) (SGDX20201103095002019), Shenzhen Basic Research Program (JCYJ20210324130209023) of Shenzhen Science and Technology Innovation Committee, NSF of Jiangsu Province (No. BK20201441), Jiangsu Post-doctoral Research Funding Program (No. 2020Z020), and NSFC (Grant No. 82072019).

\section{Conflict of Interest}
The authors reported no potential conflict of interest.
\section{Data Availability Statement} 
The data used in this study is privately owned and not publicly available. It is stored on a secure server located within our organization's premises. Access to the data is restricted to authorized personnel only, and any request for data access should be made through a formal data access agreement. 

\bibliography{ref}


\begin{thebibliography}{60}
\ifx \bisbn   \undefined \def \bisbn  #1{ISBN #1}\fi
\ifx \binits  \undefined \def \binits#1{#1}\fi
\ifx \bauthor  \undefined \def \bauthor#1{#1}\fi
\ifx \batitle  \undefined \def \batitle#1{#1}\fi
\ifx \bjtitle  \undefined \def \bjtitle#1{#1}\fi
\ifx \bvolume  \undefined \def \bvolume#1{\textbf{#1}}\fi
\ifx \byear  \undefined \def \byear#1{#1}\fi
\ifx \bissue  \undefined \def \bissue#1{#1}\fi
\ifx \bfpage  \undefined \def \bfpage#1{#1}\fi
\ifx \blpage  \undefined \def \blpage #1{#1}\fi
\ifx \burl  \undefined \def \burl#1{\textsf{#1}}\fi
\ifx \doiurl  \undefined \def \doiurl#1{\url{https://doi.org/#1}}\fi
\ifx \betal  \undefined \def \betal{\textit{et al.}}\fi
\ifx \binstitute  \undefined \def \binstitute#1{#1}\fi
\ifx \binstitutionaled  \undefined \def \binstitutionaled#1{#1}\fi
\ifx \bctitle  \undefined \def \bctitle#1{#1}\fi
\ifx \beditor  \undefined \def \beditor#1{#1}\fi
\ifx \bpublisher  \undefined \def \bpublisher#1{#1}\fi
\ifx \bbtitle  \undefined \def \bbtitle#1{#1}\fi
\ifx \bedition  \undefined \def \bedition#1{#1}\fi
\ifx \bseriesno  \undefined \def \bseriesno#1{#1}\fi
\ifx \blocation  \undefined \def \blocation#1{#1}\fi
\ifx \bsertitle  \undefined \def \bsertitle#1{#1}\fi
\ifx \bsnm \undefined \def \bsnm#1{#1}\fi
\ifx \bsuffix \undefined \def \bsuffix#1{#1}\fi
\ifx \bparticle \undefined \def \bparticle#1{#1}\fi
\ifx \barticle \undefined \def \barticle#1{#1}\fi
\bibcommenthead
\ifx \bconfdate \undefined \def \bconfdate #1{#1}\fi
\ifx \botherref \undefined \def \botherref #1{#1}\fi
\ifx \url \undefined \def \url#1{\textsf{#1}}\fi
\ifx \bchapter \undefined \def \bchapter#1{#1}\fi
\ifx \bbook \undefined \def \bbook#1{#1}\fi
\ifx \bcomment \undefined \def \bcomment#1{#1}\fi
\ifx \oauthor \undefined \def \oauthor#1{#1}\fi
\ifx \citeauthoryear \undefined \def \citeauthoryear#1{#1}\fi
\ifx \endbibitem  \undefined \def \endbibitem {}\fi
\ifx \bconflocation  \undefined \def \bconflocation#1{#1}\fi
\ifx \arxivurl  \undefined \def \arxivurl#1{\textsf{#1}}\fi
\csname PreBibitemsHook\endcsname

\bibitem[\protect\citeauthoryear{Al-Sarraf
  et~al.}{1998}]{al1998chemoradiotherapy}
\begin{barticle}
\bauthor{\bsnm{Al-Sarraf}, \binits{M.}},
\bauthor{\bsnm{LeBlanc}, \binits{M.}},
\bauthor{\bsnm{Giri}, \binits{P.}},
\bauthor{\bsnm{Fu}, \binits{K.K.}},
\bauthor{\bsnm{Cooper}, \binits{J.}},
\bauthor{\bsnm{Vuong}, \binits{T.}},
\bauthor{\bsnm{Forastiere}, \binits{A.A.}},
\bauthor{\bsnm{Adams}, \binits{G.}},
\bauthor{\bsnm{Sakr}, \binits{W.A.}},
\bauthor{\bsnm{Schuller}, \binits{D.E.}}, \betal:
\batitle{Chemoradiotherapy versus radiotherapy in patients with advanced
  nasopharyngeal cancer: phase iii randomized intergroup study 0099.}
\bjtitle{Journal of clinical oncology}
\bvolume{16}(\bissue{4}),
\bfpage{1310}--\blpage{1317}
(\byear{1998})
\end{barticle}
\endbibitem

\bibitem[\protect\citeauthoryear{Cooper et~al.}{2000}]{cooper2000improved}
\begin{barticle}
\bauthor{\bsnm{Cooper}, \binits{J.S.}},
\bauthor{\bsnm{Lee}, \binits{H.}},
\bauthor{\bsnm{Torrey}, \binits{M.}},
\bauthor{\bsnm{Hochster}, \binits{H.}}:
\batitle{Improved outcome secondary to concurrent chemoradiotherapy for
  advanced carcinoma of the nasopharynx: preliminary corroboration of the
  intergroup experience}.
\bjtitle{International Journal of Radiation Oncology* Biology* Physics}
\bvolume{47}(\bissue{4}),
\bfpage{861}--\blpage{866}
(\byear{2000})
\end{barticle}
\endbibitem

\bibitem[\protect\citeauthoryear{Liu et~al.}{2019}]{liu2019treatment}
\begin{barticle}
\bauthor{\bsnm{Liu}, \binits{Z.}},
\bauthor{\bsnm{Xu}, \binits{C.}},
\bauthor{\bsnm{Jiang}, \binits{R.}},
\bauthor{\bsnm{Liu}, \binits{G.}},
\bauthor{\bsnm{Liu}, \binits{Q.}},
\bauthor{\bsnm{Zhou}, \binits{J.}},
\bauthor{\bsnm{Liu}, \binits{F.}},
\bauthor{\bsnm{Yao}, \binits{J.}},
\bauthor{\bsnm{Wang}, \binits{S.}},
\bauthor{\bsnm{Jiang}, \binits{W.}}:
\batitle{Treatment of locally advanced nasopharyngeal carcinoma by helical
  tomotherapy: an observational, prospective analysis}.
\bjtitle{Translational oncology}
\bvolume{12}(\bissue{5}),
\bfpage{757}--\blpage{763}
(\byear{2019})
\end{barticle}
\endbibitem

\bibitem[\protect\citeauthoryear{Lin et~al.}{2012}]{lin2012combined}
\begin{barticle}
\bauthor{\bsnm{Lin}, \binits{S.}},
\bauthor{\bsnm{Tham}, \binits{I.W.}},
\bauthor{\bsnm{Pan}, \binits{J.}},
\bauthor{\bsnm{Han}, \binits{L.}},
\bauthor{\bsnm{Chen}, \binits{Q.}},
\bauthor{\bsnm{Lu}, \binits{J.J.}}:
\batitle{Combined high-dose radiation therapy and systemic chemotherapy
  improves survival in patients with newly diagnosed metastatic nasopharyngeal
  cancer}.
\bjtitle{American journal of clinical oncology}
\bvolume{35}(\bissue{5}),
\bfpage{474}--\blpage{479}
(\byear{2012})
\end{barticle}
\endbibitem

\bibitem[\protect\citeauthoryear{Marks et~al.}{1982}]{marks1982dose}
\begin{barticle}
\bauthor{\bsnm{Marks}, \binits{J.E.}},
\bauthor{\bsnm{Bedwinek}, \binits{J.M.}},
\bauthor{\bsnm{Lee}, \binits{F.}},
\bauthor{\bsnm{Purdy}, \binits{J.A.}},
\bauthor{\bsnm{Perez}, \binits{C.A.}}:
\batitle{Dose-response analysis for nasopharyngeal carcinoma. an historical
  perspective}.
\bjtitle{Cancer}
\bvolume{50}(\bissue{6}),
\bfpage{1042}--\blpage{1050}
(\byear{1982})
\end{barticle}
\endbibitem

\bibitem[\protect\citeauthoryear{Vikram et~al.}{1985}]{vikram1985patterns}
\begin{barticle}
\bauthor{\bsnm{Vikram}, \binits{B.}},
\bauthor{\bsnm{Mishra}, \binits{U.B.}},
\bauthor{\bsnm{Strong}, \binits{E.W.}},
\bauthor{\bsnm{Manolatos}, \binits{S.}}:
\batitle{Patterns of failure in carcinoma of the nasopharynx: I. failure at the
  primary site}.
\bjtitle{International Journal of Radiation Oncology* Biology* Physics}
\bvolume{11}(\bissue{8}),
\bfpage{1455}--\blpage{1459}
(\byear{1985})
\end{barticle}
\endbibitem

\bibitem[\protect\citeauthoryear{Zhao et~al.}{2018}]{zhao2018molecular}
\begin{barticle}
\bauthor{\bsnm{Zhao}, \binits{L.}},
\bauthor{\bsnm{Fong}, \binits{A.H.}},
\bauthor{\bsnm{Liu}, \binits{N.}},
\bauthor{\bsnm{Cho}, \binits{W.}}:
\batitle{Molecular subtyping of nasopharyngeal carcinoma (npc) and a
  microrna-based prognostic model for distant metastasis}.
\bjtitle{Journal of biomedical science}
\bvolume{25}(\bissue{1}),
\bfpage{1}--\blpage{12}
(\byear{2018})
\end{barticle}
\endbibitem

\bibitem[\protect\citeauthoryear{Zhang et~al.}{2019}]{zhang2019development}
\begin{barticle}
\bauthor{\bsnm{Zhang}, \binits{L.}},
\bauthor{\bsnm{Dong}, \binits{D.}},
\bauthor{\bsnm{Li}, \binits{H.}},
\bauthor{\bsnm{Tian}, \binits{J.}},
\bauthor{\bsnm{Ouyang}, \binits{F.}},
\bauthor{\bsnm{Mo}, \binits{X.}},
\bauthor{\bsnm{Zhang}, \binits{B.}},
\bauthor{\bsnm{Luo}, \binits{X.}},
\bauthor{\bsnm{Lian}, \binits{Z.}},
\bauthor{\bsnm{Pei}, \binits{S.}}, \betal:
\batitle{Development and validation of a magnetic resonance imaging-based model
  for the prediction of distant metastasis before initial treatment of
  nasopharyngeal carcinoma: a retrospective cohort study}.
\bjtitle{EBioMedicine}
\bvolume{40},
\bfpage{327}--\blpage{335}
(\byear{2019})
\end{barticle}
\endbibitem

\bibitem[\protect\citeauthoryear{Lam et~al.}{2022}]{lam2022multi}
\begin{botherref}
\oauthor{\bsnm{Lam}, \binits{S.K.}}, et al.:
Multi-organ multi-omics prediction of adaptive radiotherapy eligibility in
  patients with nasopharyngeal carcinoma
(2022)
\end{botherref}
\endbibitem

\bibitem[\protect\citeauthoryear{Zhang et~al.}{2023}]{zhang2023radiomic}
\begin{barticle}
\bauthor{\bsnm{Zhang}, \binits{J.}},
\bauthor{\bsnm{Lam}, \binits{S.-K.}},
\bauthor{\bsnm{Teng}, \binits{X.}},
\bauthor{\bsnm{Ma}, \binits{Z.}},
\bauthor{\bsnm{Han}, \binits{X.}},
\bauthor{\bsnm{Zhang}, \binits{Y.}},
\bauthor{\bsnm{Cheung}, \binits{A.L.-Y.}},
\bauthor{\bsnm{Chau}, \binits{T.-C.}},
\bauthor{\bsnm{Ng}, \binits{S.C.-Y.}},
\bauthor{\bsnm{Lee}, \binits{F.K.-H.}}, \betal:
\batitle{Radiomic feature repeatability and its impact on prognostic model
  generalizability: A multi-institutional study on nasopharyngeal carcinoma
  patients}.
\bjtitle{Radiotherapy and Oncology}
\bvolume{183},
\bfpage{109578}
(\byear{2023})
\end{barticle}
\endbibitem

\bibitem[\protect\citeauthoryear{Zhang et~al.}{2022}]{zhang2022integration}
\begin{barticle}
\bauthor{\bsnm{Zhang}, \binits{Y.}},
\bauthor{\bsnm{Lam}, \binits{S.}},
\bauthor{\bsnm{Yu}, \binits{T.}},
\bauthor{\bsnm{Teng}, \binits{X.}},
\bauthor{\bsnm{Zhang}, \binits{J.}},
\bauthor{\bsnm{Lee}, \binits{F.K.-h.}},
\bauthor{\bsnm{Au}, \binits{K.-h.}},
\bauthor{\bsnm{Yip}, \binits{C.W.-y.}},
\bauthor{\bsnm{Wang}, \binits{S.}},
\bauthor{\bsnm{Cai}, \binits{J.}}:
\batitle{Integration of an imbalance framework with novel high-generalizable
  classifiers for radiomics-based distant metastases prediction of advanced
  nasopharyngeal carcinoma}.
\bjtitle{Knowledge-based systems}
\bvolume{235},
\bfpage{107649}
(\byear{2022})
\end{barticle}
\endbibitem

\bibitem[\protect\citeauthoryear{Chen et~al.}{2013}]{chen2013efficient}
\begin{barticle}
\bauthor{\bsnm{Chen}, \binits{H.-L.}},
\bauthor{\bsnm{Huang}, \binits{C.-C.}},
\bauthor{\bsnm{Yu}, \binits{X.-G.}},
\bauthor{\bsnm{Xu}, \binits{X.}},
\bauthor{\bsnm{Sun}, \binits{X.}},
\bauthor{\bsnm{Wang}, \binits{G.}},
\bauthor{\bsnm{Wang}, \binits{S.-J.}}:
\batitle{An efficient diagnosis system for detection of parkinson’s disease
  using fuzzy k-nearest neighbor approach}.
\bjtitle{Expert systems with applications}
\bvolume{40}(\bissue{1}),
\bfpage{263}--\blpage{271}
(\byear{2013})
\end{barticle}
\endbibitem

\bibitem[\protect\citeauthoryear{Chen et~al.}{2016}]{chen2016efficient}
\begin{barticle}
\bauthor{\bsnm{Chen}, \binits{H.-L.}},
\bauthor{\bsnm{Wang}, \binits{G.}},
\bauthor{\bsnm{Ma}, \binits{C.}},
\bauthor{\bsnm{Cai}, \binits{Z.-N.}},
\bauthor{\bsnm{Liu}, \binits{W.-B.}},
\bauthor{\bsnm{Wang}, \binits{S.-J.}}:
\batitle{An efficient hybrid kernel extreme learning machine approach for early
  diagnosis of parkinson's disease}.
\bjtitle{Neurocomputing}
\bvolume{184},
\bfpage{131}--\blpage{144}
(\byear{2016})
\end{barticle}
\endbibitem

\bibitem[\protect\citeauthoryear{Chen et~al.}{2012}]{chen2012support}
\begin{barticle}
\bauthor{\bsnm{Chen}, \binits{H.-L.}},
\bauthor{\bsnm{Yang}, \binits{B.}},
\bauthor{\bsnm{Wang}, \binits{G.}},
\bauthor{\bsnm{Wang}, \binits{S.-J.}},
\bauthor{\bsnm{Liu}, \binits{J.}},
\bauthor{\bsnm{Liu}, \binits{D.-Y.}}:
\batitle{Support vector machine based diagnostic system for breast cancer using
  swarm intelligence}.
\bjtitle{Journal of medical systems}
\bvolume{36},
\bfpage{2505}--\blpage{2519}
(\byear{2012})
\end{barticle}
\endbibitem

\bibitem[\protect\citeauthoryear{Zhang et~al.}{2022}]{zhang2022repeatability}
\begin{bchapter}
\bauthor{\bsnm{Zhang}, \binits{J.}},
\bauthor{\bsnm{Lam}, \binits{S.}},
\bauthor{\bsnm{Teng}, \binits{X.}},
\bauthor{\bsnm{Zhang}, \binits{Y.}},
\bauthor{\bsnm{Ma}, \binits{Z.}},
\bauthor{\bsnm{Lee}, \binits{F.}},
\bauthor{\bsnm{Au}, \binits{K.-h.}},
\bauthor{\bsnm{Yip}, \binits{W.Y.}},
\bauthor{\bsnm{Chang}, \binits{T.Y.A.}},
\bauthor{\bsnm{Chan}, \binits{W.C.L.}}, \betal:
\bctitle{Repeatability of radiomic features against simulated scanning position
  stochasticity across imaging modalities and cancer subtypes: A retrospective
  multi-institutional study on head-and-neck cases}.
In: \bbtitle{Computational Mathematics Modeling in Cancer Analysis: First
  International Workshop, CMMCA 2022, Held in Conjunction with MICCAI 2022,
  Singapore, September 18, 2022, Proceedings},
pp. \bfpage{21}--\blpage{34}
(\byear{2022}).
\bcomment{Springer}
\end{bchapter}
\endbibitem

\bibitem[\protect\citeauthoryear{Teng et~al.}{2022a}]{teng2022improving}
\begin{botherref}
\oauthor{\bsnm{Teng}, \binits{X.}},
\oauthor{\bsnm{Zhang}, \binits{J.}},
\oauthor{\bsnm{Ma}, \binits{Z.}},
\oauthor{\bsnm{Zhang}, \binits{Y.}},
\oauthor{\bsnm{Lam}, \binits{S.}},
\oauthor{\bsnm{Li}, \binits{W.}},
\oauthor{\bsnm{Xiao}, \binits{H.}},
\oauthor{\bsnm{Li}, \binits{T.}},
\oauthor{\bsnm{Li}, \binits{B.}},
\oauthor{\bsnm{Zhou}, \binits{T.}}, et al.:
Improving radiomic model reliability using robust features from perturbations
  for head-and-neck carcinoma.
Frontiers in Oncology
\textbf{12}
(2022)
\end{botherref}
\endbibitem

\bibitem[\protect\citeauthoryear{Teng et~al.}{2022b}]{teng2022building}
\begin{barticle}
\bauthor{\bsnm{Teng}, \binits{X.}},
\bauthor{\bsnm{Zhang}, \binits{J.}},
\bauthor{\bsnm{Zwanenburg}, \binits{A.}},
\bauthor{\bsnm{Sun}, \binits{J.}},
\bauthor{\bsnm{Huang}, \binits{Y.}},
\bauthor{\bsnm{Lam}, \binits{S.}},
\bauthor{\bsnm{Zhang}, \binits{Y.}},
\bauthor{\bsnm{Li}, \binits{B.}},
\bauthor{\bsnm{Zhou}, \binits{T.}},
\bauthor{\bsnm{Xiao}, \binits{H.}}, \betal:
\batitle{Building reliable radiomic models using image perturbation}.
\bjtitle{Scientific Reports}
\bvolume{12}(\bissue{1}),
\bfpage{1}--\blpage{10}
(\byear{2022})
\end{barticle}
\endbibitem

\bibitem[\protect\citeauthoryear{Chakraborty
  et~al.}{2018}]{chakraborty2018onco}
\begin{botherref}
\oauthor{\bsnm{Chakraborty}, \binits{S.}},
\oauthor{\bsnm{Hosen}, \binits{M.I.}},
\oauthor{\bsnm{Ahmed}, \binits{M.}},
\oauthor{\bsnm{Shekhar}, \binits{H.U.}}:
Onco-multi-omics approach: a new frontier in cancer research.
BioMed research international
\textbf{2018}
(2018)
\end{botherref}
\endbibitem

\bibitem[\protect\citeauthoryear{Olivier et~al.}{2019}]{olivier2019need}
\begin{barticle}
\bauthor{\bsnm{Olivier}, \binits{M.}},
\bauthor{\bsnm{Asmis}, \binits{R.}},
\bauthor{\bsnm{Hawkins}, \binits{G.A.}},
\bauthor{\bsnm{Howard}, \binits{T.D.}},
\bauthor{\bsnm{Cox}, \binits{L.A.}}:
\batitle{The need for multi-omics biomarker signatures in precision medicine}.
\bjtitle{International journal of molecular sciences}
\bvolume{20}(\bissue{19}),
\bfpage{4781}
(\byear{2019})
\end{barticle}
\endbibitem

\bibitem[\protect\citeauthoryear{Hu and Jia}{2021}]{hu2021multi}
\begin{barticle}
\bauthor{\bsnm{Hu}, \binits{C.}},
\bauthor{\bsnm{Jia}, \binits{W.}}:
\batitle{Multi-omics profiling: the way toward precision medicine in metabolic
  diseases}.
\bjtitle{Journal of Molecular Cell Biology}
\bvolume{13}(\bissue{8}),
\bfpage{576}--\blpage{593}
(\byear{2021})
\end{barticle}
\endbibitem

\bibitem[\protect\citeauthoryear{Zhu et~al.}{2022}]{zhu2022deep}
\begin{botherref}
\oauthor{\bsnm{Zhu}, \binits{Q.}},
\oauthor{\bsnm{Xu}, \binits{B.}},
\oauthor{\bsnm{Huang}, \binits{J.}},
\oauthor{\bsnm{Wang}, \binits{H.}},
\oauthor{\bsnm{Xu}, \binits{R.}},
\oauthor{\bsnm{Shao}, \binits{W.}},
\oauthor{\bsnm{Zhang}, \binits{D.}}:
Deep multi-modal discriminative and interpretability network for alzheimer’s
  disease diagnosis.
IEEE Transactions on Medical Imaging
(2022)
\end{botherref}
\endbibitem

\bibitem[\protect\citeauthoryear{Zhou et~al.}{2019}]{zhou2019latent}
\begin{barticle}
\bauthor{\bsnm{Zhou}, \binits{T.}},
\bauthor{\bsnm{Liu}, \binits{M.}},
\bauthor{\bsnm{Thung}, \binits{K.-H.}},
\bauthor{\bsnm{Shen}, \binits{D.}}:
\batitle{Latent representation learning for alzheimer’s disease diagnosis
  with incomplete multi-modality neuroimaging and genetic data}.
\bjtitle{IEEE transactions on medical imaging}
\bvolume{38}(\bissue{10}),
\bfpage{2411}--\blpage{2422}
(\byear{2019})
\end{barticle}
\endbibitem

\bibitem[\protect\citeauthoryear{Dong et~al.}{2023}]{dong2023multimodal}
\begin{barticle}
\bauthor{\bsnm{Dong}, \binits{Y.}},
\bauthor{\bsnm{Zhang}, \binits{J.}},
\bauthor{\bsnm{Lam}, \binits{S.}},
\bauthor{\bsnm{Zhang}, \binits{X.}},
\bauthor{\bsnm{Liu}, \binits{A.}},
\bauthor{\bsnm{Teng}, \binits{X.}},
\bauthor{\bsnm{Han}, \binits{X.}},
\bauthor{\bsnm{Cao}, \binits{J.}},
\bauthor{\bsnm{Li}, \binits{H.}},
\bauthor{\bsnm{Lee}, \binits{F.K.}}, \betal:
\batitle{Multimodal data integration to predict severe acute oral mucositis of
  nasopharyngeal carcinoma patients following radiation therapy}.
\bjtitle{Cancers}
\bvolume{15}(\bissue{7}),
\bfpage{2032}
(\byear{2023})
\end{barticle}
\endbibitem

\bibitem[\protect\citeauthoryear{Ho et~al.}{2023}]{ho2023association}
\begin{barticle}
\bauthor{\bsnm{Ho}, \binits{L.-M.}},
\bauthor{\bsnm{Lam}, \binits{S.-K.}},
\bauthor{\bsnm{Zhang}, \binits{J.}},
\bauthor{\bsnm{Chiang}, \binits{C.-L.}},
\bauthor{\bsnm{Chan}, \binits{A.C.-Y.}},
\bauthor{\bsnm{Cai}, \binits{J.}}:
\batitle{Association of multi-phasic mr-based radiomic and dosimetric features
  with treatment response in unresectable hepatocellular carcinoma patients
  following novel sequential tace-sbrt-immunotherapy}.
\bjtitle{Cancers}
\bvolume{15}(\bissue{4}),
\bfpage{1105}
(\byear{2023})
\end{barticle}
\endbibitem

\bibitem[\protect\citeauthoryear{Li et~al.}{2022}]{li2022function}
\begin{botherref}
\oauthor{\bsnm{Li}, \binits{B.}},
\oauthor{\bsnm{Ren}, \binits{G.}},
\oauthor{\bsnm{Guo}, \binits{W.}},
\oauthor{\bsnm{Zhang}, \binits{J.}},
\oauthor{\bsnm{Lam}, \binits{S.-K.}},
\oauthor{\bsnm{Zheng}, \binits{X.}},
\oauthor{\bsnm{Teng}, \binits{X.}},
\oauthor{\bsnm{Wang}, \binits{Y.}},
\oauthor{\bsnm{Yang}, \binits{Y.}},
\oauthor{\bsnm{Dan}, \binits{Q.}}, et al.:
Function-wise dual-omics analysis for radiation pneumonitis prediction in lung
  cancer patients.
Computational Intelligence in Personalized Medicine,
110
(2022)
\end{botherref}
\endbibitem

\bibitem[\protect\citeauthoryear{Zheng et~al.}{2023}]{zheng2023multi}
\begin{barticle}
\bauthor{\bsnm{Zheng}, \binits{X.}},
\bauthor{\bsnm{Guo}, \binits{W.}},
\bauthor{\bsnm{Wang}, \binits{Y.}},
\bauthor{\bsnm{Zhang}, \binits{J.}},
\bauthor{\bsnm{Zhang}, \binits{Y.}},
\bauthor{\bsnm{Cheng}, \binits{C.}},
\bauthor{\bsnm{Teng}, \binits{X.}},
\bauthor{\bsnm{Lam}, \binits{S.}},
\bauthor{\bsnm{Zhou}, \binits{T.}},
\bauthor{\bsnm{Ma}, \binits{Z.}}, \betal:
\batitle{Multi-omics to predict acute radiation esophagitis in patients with
  lung cancer treated with intensity-modulated radiation therapy}.
\bjtitle{European Journal of Medical Research}
\bvolume{28}(\bissue{1}),
\bfpage{1}--\blpage{10}
(\byear{2023})
\end{barticle}
\endbibitem

\bibitem[\protect\citeauthoryear{Chen et~al.}{2011a}]{chen2011novel}
\begin{barticle}
\bauthor{\bsnm{Chen}, \binits{H.-L.}},
\bauthor{\bsnm{Yang}, \binits{B.}},
\bauthor{\bsnm{Wang}, \binits{G.}},
\bauthor{\bsnm{Liu}, \binits{J.}},
\bauthor{\bsnm{Xu}, \binits{X.}},
\bauthor{\bsnm{Wang}, \binits{S.-J.}},
\bauthor{\bsnm{Liu}, \binits{D.-Y.}}:
\batitle{A novel bankruptcy prediction model based on an adaptive fuzzy
  k-nearest neighbor method}.
\bjtitle{Knowledge-Based Systems}
\bvolume{24}(\bissue{8}),
\bfpage{1348}--\blpage{1359}
(\byear{2011})
\end{barticle}
\endbibitem

\bibitem[\protect\citeauthoryear{Chen et~al.}{2011b}]{chen2011support}
\begin{barticle}
\bauthor{\bsnm{Chen}, \binits{H.-L.}},
\bauthor{\bsnm{Yang}, \binits{B.}},
\bauthor{\bsnm{Liu}, \binits{J.}},
\bauthor{\bsnm{Liu}, \binits{D.-Y.}}:
\batitle{A support vector machine classifier with rough set-based feature
  selection for breast cancer diagnosis}.
\bjtitle{Expert systems with applications}
\bvolume{38}(\bissue{7}),
\bfpage{9014}--\blpage{9022}
(\byear{2011})
\end{barticle}
\endbibitem

\bibitem[\protect\citeauthoryear{Zuo et~al.}{2013}]{zuo2013effective}
\begin{barticle}
\bauthor{\bsnm{Zuo}, \binits{W.-L.}},
\bauthor{\bsnm{Wang}, \binits{Z.-Y.}},
\bauthor{\bsnm{Liu}, \binits{T.}},
\bauthor{\bsnm{Chen}, \binits{H.-L.}}:
\batitle{Effective detection of parkinson's disease using an adaptive fuzzy
  k-nearest neighbor approach}.
\bjtitle{Biomedical Signal Processing and Control}
\bvolume{8}(\bissue{4}),
\bfpage{364}--\blpage{373}
(\byear{2013})
\end{barticle}
\endbibitem

\bibitem[\protect\citeauthoryear{Jing et~al.}{2020}]{jing2020adaptive}
\begin{barticle}
\bauthor{\bsnm{Jing}, \binits{M.}},
\bauthor{\bsnm{Zhao}, \binits{J.}},
\bauthor{\bsnm{Li}, \binits{J.}},
\bauthor{\bsnm{Zhu}, \binits{L.}},
\bauthor{\bsnm{Yang}, \binits{Y.}},
\bauthor{\bsnm{Shen}, \binits{H.T.}}:
\batitle{Adaptive component embedding for domain adaptation}.
\bjtitle{IEEE transactions on cybernetics}
\bvolume{51}(\bissue{7}),
\bfpage{3390}--\blpage{3403}
(\byear{2020})
\end{barticle}
\endbibitem

\bibitem[\protect\citeauthoryear{Tian and Feng}{2021}]{tian2021large}
\begin{barticle}
\bauthor{\bsnm{Tian}, \binits{Y.}},
\bauthor{\bsnm{Feng}, \binits{X.}}:
\batitle{Large margin graph embedding-based discriminant dimensionality
  reduction}.
\bjtitle{Scientific Programming}
\bvolume{2021},
\bfpage{1}--\blpage{12}
(\byear{2021})
\end{barticle}
\endbibitem

\bibitem[\protect\citeauthoryear{Kang et~al.}{2021}]{kang2021structured}
\begin{barticle}
\bauthor{\bsnm{Kang}, \binits{Z.}},
\bauthor{\bsnm{Peng}, \binits{C.}},
\bauthor{\bsnm{Cheng}, \binits{Q.}},
\bauthor{\bsnm{Liu}, \binits{X.}},
\bauthor{\bsnm{Peng}, \binits{X.}},
\bauthor{\bsnm{Xu}, \binits{Z.}},
\bauthor{\bsnm{Tian}, \binits{L.}}:
\batitle{Structured graph learning for clustering and semi-supervised
  classification}.
\bjtitle{Pattern Recognition}
\bvolume{110},
\bfpage{107627}
(\byear{2021})
\end{barticle}
\endbibitem

\bibitem[\protect\citeauthoryear{Song et~al.}{2022}]{song2022graph}
\begin{botherref}
\oauthor{\bsnm{Song}, \binits{Z.}},
\oauthor{\bsnm{Yang}, \binits{X.}},
\oauthor{\bsnm{Xu}, \binits{Z.}},
\oauthor{\bsnm{King}, \binits{I.}}:
Graph-based semi-supervised learning: A comprehensive review.
IEEE Transactions on Neural Networks and Learning Systems
(2022)
\end{botherref}
\endbibitem

\bibitem[\protect\citeauthoryear{G{\"o}nen and
  Alpayd{\i}n}{2011}]{gonen2011multiple}
\begin{barticle}
\bauthor{\bsnm{G{\"o}nen}, \binits{M.}},
\bauthor{\bsnm{Alpayd{\i}n}, \binits{E.}}:
\batitle{Multiple kernel learning algorithms}.
\bjtitle{The Journal of Machine Learning Research}
\bvolume{12},
\bfpage{2211}--\blpage{2268}
(\byear{2011})
\end{barticle}
\endbibitem

\bibitem[\protect\citeauthoryear{Bucak et~al.}{2013}]{bucak2013multiple}
\begin{barticle}
\bauthor{\bsnm{Bucak}, \binits{S.S.}},
\bauthor{\bsnm{Jin}, \binits{R.}},
\bauthor{\bsnm{Jain}, \binits{A.K.}}:
\batitle{Multiple kernel learning for visual object recognition: A review}.
\bjtitle{IEEE Transactions on Pattern Analysis and Machine Intelligence}
\bvolume{36}(\bissue{7}),
\bfpage{1354}--\blpage{1369}
(\byear{2013})
\end{barticle}
\endbibitem

\bibitem[\protect\citeauthoryear{Niazmardi
  et~al.}{2017}]{niazmardi2017multiple}
\begin{barticle}
\bauthor{\bsnm{Niazmardi}, \binits{S.}},
\bauthor{\bsnm{Demir}, \binits{B.}},
\bauthor{\bsnm{Bruzzone}, \binits{L.}},
\bauthor{\bsnm{Safari}, \binits{A.}},
\bauthor{\bsnm{Homayouni}, \binits{S.}}:
\batitle{Multiple kernel learning for remote sensing image classification}.
\bjtitle{IEEE Transactions on Geoscience and Remote Sensing}
\bvolume{56}(\bissue{3}),
\bfpage{1425}--\blpage{1443}
(\byear{2017})
\end{barticle}
\endbibitem

\bibitem[\protect\citeauthoryear{Aiolli and Donini}{2015}]{aiolli2015easymkl}
\begin{barticle}
\bauthor{\bsnm{Aiolli}, \binits{F.}},
\bauthor{\bsnm{Donini}, \binits{M.}}:
\batitle{Easymkl: a scalable multiple kernel learning algorithm}.
\bjtitle{Neurocomputing}
\bvolume{169},
\bfpage{215}--\blpage{224}
(\byear{2015})
\end{barticle}
\endbibitem

\bibitem[\protect\citeauthoryear{Alioscha-Perez
  et~al.}{2019}]{alioscha2019svrg}
\begin{barticle}
\bauthor{\bsnm{Alioscha-Perez}, \binits{M.}},
\bauthor{\bsnm{Oveneke}, \binits{M.C.}},
\bauthor{\bsnm{Sahli}, \binits{H.}}:
\batitle{Svrg-mkl: a fast and scalable multiple kernel learning solution for
  features combination in multi-class classification problems}.
\bjtitle{IEEE transactions on neural networks and learning systems}
\bvolume{31}(\bissue{5}),
\bfpage{1710}--\blpage{1723}
(\byear{2019})
\end{barticle}
\endbibitem

\bibitem[\protect\citeauthoryear{Xu et~al.}{2010}]{xu2010smooth}
\begin{bchapter}
\bauthor{\bsnm{Xu}, \binits{Z.}},
\bauthor{\bsnm{Jin}, \binits{R.}},
\bauthor{\bsnm{Zhu}, \binits{S.}},
\bauthor{\bsnm{Lyu}, \binits{M.}},
\bauthor{\bsnm{King}, \binits{I.}}:
\bctitle{Smooth optimization for effective multiple kernel learning}.
In: \bbtitle{Proceedings of the AAAI Conference on Artificial Intelligence},
vol. \bseriesno{24},
pp. \bfpage{637}--\blpage{642}
(\byear{2010})
\end{bchapter}
\endbibitem

\bibitem[\protect\citeauthoryear{Varma and Babu}{2009}]{varma2009more}
\begin{bchapter}
\bauthor{\bsnm{Varma}, \binits{M.}},
\bauthor{\bsnm{Babu}, \binits{B.R.}}:
\bctitle{More generality in efficient multiple kernel learning}.
In: \bbtitle{Proceedings of the 26th Annual International Conference on Machine
  Learning},
pp. \bfpage{1065}--\blpage{1072}
(\byear{2009})
\end{bchapter}
\endbibitem

\bibitem[\protect\citeauthoryear{Cortes et~al.}{2009}]{cortes2009learning}
\begin{botherref}
\oauthor{\bsnm{Cortes}, \binits{C.}},
\oauthor{\bsnm{Mohri}, \binits{M.}},
\oauthor{\bsnm{Rostamizadeh}, \binits{A.}}:
Learning non-linear combinations of kernels.
Advances in neural information processing systems
\textbf{22}
(2009)
\end{botherref}
\endbibitem

\bibitem[\protect\citeauthoryear{Zhou et~al.}{2016}]{zhou2016veto}
\begin{bchapter}
\bauthor{\bsnm{Zhou}, \binits{Y.}},
\bauthor{\bsnm{Hu}, \binits{N.}},
\bauthor{\bsnm{Spanos}, \binits{C.J.}}:
\bctitle{Veto-consensus multiple kernel learning}.
In: \bbtitle{Thirtieth AAAI Conference on Artificial Intelligence}
(\byear{2016})
\end{bchapter}
\endbibitem

\bibitem[\protect\citeauthoryear{Han et~al.}{2013}]{han2013localized}
\begin{barticle}
\bauthor{\bsnm{Han}, \binits{Y.}},
\bauthor{\bsnm{Yang}, \binits{K.}},
\bauthor{\bsnm{Ma}, \binits{Y.}},
\bauthor{\bsnm{Liu}, \binits{G.}}:
\batitle{Localized multiple kernel learning via sample-wise alternating
  optimization}.
\bjtitle{IEEE transactions on cybernetics}
\bvolume{44}(\bissue{1}),
\bfpage{137}--\blpage{148}
(\byear{2013})
\end{barticle}
\endbibitem

\bibitem[\protect\citeauthoryear{Rakotomamonjy
  et~al.}{2008}]{rakotomamonjy2008simplemkl}
\begin{barticle}
\bauthor{\bsnm{Rakotomamonjy}, \binits{A.}},
\bauthor{\bsnm{Bach}, \binits{F.}},
\bauthor{\bsnm{Canu}, \binits{S.}}, \betal:
\batitle{Simplemkl}.
\bjtitle{Journal of Machine Learning Research}
\bvolume{9},
\bfpage{2491}--\blpage{2521}
(\byear{2008})
\end{barticle}
\endbibitem

\bibitem[\protect\citeauthoryear{Li et~al.}{2008}]{li2008discriminant}
\begin{barticle}
\bauthor{\bsnm{Li}, \binits{X.}},
\bauthor{\bsnm{Lin}, \binits{S.}},
\bauthor{\bsnm{Yan}, \binits{S.}},
\bauthor{\bsnm{Xu}, \binits{D.}}:
\batitle{Discriminant locally linear embedding with high-order tensor data}.
\bjtitle{IEEE Transactions on Systems, Man, and Cybernetics, Part B
  (Cybernetics)}
\bvolume{38}(\bissue{2}),
\bfpage{342}--\blpage{352}
(\byear{2008})
\end{barticle}
\endbibitem

\bibitem[\protect\citeauthoryear{Yan et~al.}{2006}]{yan2006graph}
\begin{barticle}
\bauthor{\bsnm{Yan}, \binits{S.}},
\bauthor{\bsnm{Xu}, \binits{D.}},
\bauthor{\bsnm{Zhang}, \binits{B.}},
\bauthor{\bsnm{Zhang}, \binits{H.-J.}},
\bauthor{\bsnm{Yang}, \binits{Q.}},
\bauthor{\bsnm{Lin}, \binits{S.}}:
\batitle{Graph embedding and extensions: A general framework for dimensionality
  reduction}.
\bjtitle{IEEE transactions on pattern analysis and machine intelligence}
\bvolume{29}(\bissue{1}),
\bfpage{40}--\blpage{51}
(\byear{2006})
\end{barticle}
\endbibitem

\bibitem[\protect\citeauthoryear{Fan et~al.}{2011}]{fan2011local}
\begin{barticle}
\bauthor{\bsnm{Fan}, \binits{Z.}},
\bauthor{\bsnm{Xu}, \binits{Y.}},
\bauthor{\bsnm{Zhang}, \binits{D.}}:
\batitle{Local linear discriminant analysis framework using sample neighbors}.
\bjtitle{IEEE Transactions on Neural Networks}
\bvolume{22}(\bissue{7}),
\bfpage{1119}--\blpage{1132}
(\byear{2011})
\end{barticle}
\endbibitem

\bibitem[\protect\citeauthoryear{Sugiyama}{2007}]{sugiyama2007dimensionality}
\begin{botherref}
\oauthor{\bsnm{Sugiyama}, \binits{M.}}:
Dimensionality reduction of multimodal labeled data by local fisher
  discriminant analysis.
Journal of machine learning research
\textbf{8}(5)
(2007)
\end{botherref}
\endbibitem

\bibitem[\protect\citeauthoryear{Cai et~al.}{2007}]{cai2007locality}
\begin{bchapter}
\bauthor{\bsnm{Cai}, \binits{D.}},
\bauthor{\bsnm{He}, \binits{X.}},
\bauthor{\bsnm{Zhou}, \binits{K.}},
\bauthor{\bsnm{Han}, \binits{J.}},
\bauthor{\bsnm{Bao}, \binits{H.}}:
\bctitle{Locality sensitive discriminant analysis.}
In: \bbtitle{IJCAI},
vol. \bseriesno{2007},
pp. \bfpage{1713}--\blpage{1726}
(\byear{2007})
\end{bchapter}
\endbibitem

\bibitem[\protect\citeauthoryear{Chen et~al.}{2005}]{chen2005local}
\begin{bchapter}
\bauthor{\bsnm{Chen}, \binits{H.-T.}},
\bauthor{\bsnm{Chang}, \binits{H.-W.}},
\bauthor{\bsnm{Liu}, \binits{T.-L.}}:
\bctitle{Local discriminant embedding and its variants}.
In: \bbtitle{2005 IEEE Computer Society Conference on Computer Vision and
  Pattern Recognition (CVPR'05)},
vol. \bseriesno{2},
pp. \bfpage{846}--\blpage{853}
(\byear{2005}).
\bcomment{IEEE}
\end{bchapter}
\endbibitem

\bibitem[\protect\citeauthoryear{Nie et~al.}{2013}]{nie2013adaptive}
\begin{bchapter}
\bauthor{\bsnm{Nie}, \binits{F.}},
\bauthor{\bsnm{Wang}, \binits{H.}},
\bauthor{\bsnm{Huang}, \binits{H.}},
\bauthor{\bsnm{Ding}, \binits{C.}}:
\bctitle{Adaptive loss minimization for semi-supervised elastic embedding}.
In: \bbtitle{Twenty-Third International Joint Conference on Artificial
  Intelligence}
(\byear{2013})
\end{bchapter}
\endbibitem

\bibitem[\protect\citeauthoryear{Belkin et~al.}{2006}]{belkin2006manifold}
\begin{botherref}
\oauthor{\bsnm{Belkin}, \binits{M.}},
\oauthor{\bsnm{Niyogi}, \binits{P.}},
\oauthor{\bsnm{Sindhwani}, \binits{V.}}:
Manifold regularization: A geometric framework for learning from labeled and
  unlabeled examples.
Journal of machine learning research
\textbf{7}(11)
(2006)
\end{botherref}
\endbibitem

\bibitem[\protect\citeauthoryear{Zwanenburg et~al.}{2020}]{zwanenburg2020image}
\begin{barticle}
\bauthor{\bsnm{Zwanenburg}, \binits{A.}},
\bauthor{\bsnm{Valli{\`e}res}, \binits{M.}},
\bauthor{\bsnm{Abdalah}, \binits{M.A.}},
\bauthor{\bsnm{Aerts}, \binits{H.J.}},
\bauthor{\bsnm{Andrearczyk}, \binits{V.}},
\bauthor{\bsnm{Apte}, \binits{A.}},
\bauthor{\bsnm{Ashrafinia}, \binits{S.}},
\bauthor{\bsnm{Bakas}, \binits{S.}},
\bauthor{\bsnm{Beukinga}, \binits{R.J.}},
\bauthor{\bsnm{Boellaard}, \binits{R.}}, \betal:
\batitle{The image biomarker standardization initiative: standardized
  quantitative radiomics for high-throughput image-based phenotyping}.
\bjtitle{Radiology}
\bvolume{295}(\bissue{2}),
\bfpage{328}--\blpage{338}
(\byear{2020})
\end{barticle}
\endbibitem

\bibitem[\protect\citeauthoryear{Bommert et~al.}{2020}]{bommert2020benchmark}
\begin{barticle}
\bauthor{\bsnm{Bommert}, \binits{A.}},
\bauthor{\bsnm{Sun}, \binits{X.}},
\bauthor{\bsnm{Bischl}, \binits{B.}},
\bauthor{\bsnm{Rahnenf{\"u}hrer}, \binits{J.}},
\bauthor{\bsnm{Lang}, \binits{M.}}:
\batitle{Benchmark for filter methods for feature selection in high-dimensional
  classification data}.
\bjtitle{Computational Statistics \& Data Analysis}
\bvolume{143},
\bfpage{106839}
(\byear{2020})
\end{barticle}
\endbibitem

\bibitem[\protect\citeauthoryear{Atif et~al.}{2022}]{atif2022multi}
\begin{botherref}
\oauthor{\bsnm{Atif}, \binits{S.M.}},
\oauthor{\bsnm{Khan}, \binits{S.}},
\oauthor{\bsnm{Naseem}, \binits{I.}},
\oauthor{\bsnm{Togneri}, \binits{R.}},
\oauthor{\bsnm{Bennamoun}, \binits{M.}}:
Multi-kernel fusion for rbf neural networks.
Neural Processing Letters,
1--25
(2022)
\end{botherref}
\endbibitem

\bibitem[\protect\citeauthoryear{Fang et~al.}{2017}]{fang2017regularized}
\begin{barticle}
\bauthor{\bsnm{Fang}, \binits{X.}},
\bauthor{\bsnm{Xu}, \binits{Y.}},
\bauthor{\bsnm{Li}, \binits{X.}},
\bauthor{\bsnm{Lai}, \binits{Z.}},
\bauthor{\bsnm{Wong}, \binits{W.K.}},
\bauthor{\bsnm{Fang}, \binits{B.}}:
\batitle{Regularized label relaxation linear regression}.
\bjtitle{IEEE transactions on neural networks and learning systems}
\bvolume{29}(\bissue{4}),
\bfpage{1006}--\blpage{1018}
(\byear{2017})
\end{barticle}
\endbibitem

\bibitem[\protect\citeauthoryear{Lin and Wang}{2002}]{lin2002fuzzy}
\begin{barticle}
\bauthor{\bsnm{Lin}, \binits{C.-F.}},
\bauthor{\bsnm{Wang}, \binits{S.-D.}}:
\batitle{Fuzzy support vector machines}.
\bjtitle{IEEE transactions on neural networks}
\bvolume{13}(\bissue{2}),
\bfpage{464}--\blpage{471}
(\byear{2002})
\end{barticle}
\endbibitem

\bibitem[\protect\citeauthoryear{Batuwita and Palade}{2010}]{batuwita2010fsvm}
\begin{barticle}
\bauthor{\bsnm{Batuwita}, \binits{R.}},
\bauthor{\bsnm{Palade}, \binits{V.}}:
\batitle{Fsvm-cil: fuzzy support vector machines for class imbalance learning}.
\bjtitle{IEEE Transactions on Fuzzy Systems}
\bvolume{18}(\bissue{3}),
\bfpage{558}--\blpage{571}
(\byear{2010})
\end{barticle}
\endbibitem

\bibitem[\protect\citeauthoryear{McDonald}{2009}]{mcdonald2009ridge}
\begin{barticle}
\bauthor{\bsnm{McDonald}, \binits{G.C.}}:
\batitle{Ridge regression}.
\bjtitle{Wiley Interdisciplinary Reviews: Computational Statistics}
\bvolume{1}(\bissue{1}),
\bfpage{93}--\blpage{100}
(\byear{2009})
\end{barticle}
\endbibitem

\bibitem[\protect\citeauthoryear{Kurita}{2019}]{kurita2019principal}
\begin{botherref}
\oauthor{\bsnm{Kurita}, \binits{T.}}:
Principal component analysis (pca).
Computer Vision: A Reference Guide,
1--4
(2019)
\end{botherref}
\endbibitem

\end{thebibliography}

\end{document}